\documentclass[12pt]{article}

\usepackage{graphics}
\usepackage{amsmath}
\usepackage{amsfonts}
\usepackage{amssymb}
\usepackage{graphicx}
\usepackage{rotating}

\begin{document}

\begin{titlepage}

\centerline{\bf CORRELATION ENERGY OF AN ELECTRON GAS: }
\vspace*{0.035truein}
\centerline{\bf A FUNCTIONAL APPROACH}
\vspace*{0.37truein}
\centerline{\footnotesize A. REBEI \footnote{Current Address: Seagate Research, Pittsburgh, PA 15222, U.S.A. } }
\vspace*{0.015truein}
\centerline{\footnotesize\it Department of Physics} 
\baselineskip=10pt
\centerline{\footnotesize\it University of Wisconsin-Madison, Madison, WI 
53706, U.S.A.}
\vspace*{10pt}
\centerline{\normalsize and}
\vspace*{10pt}
\centerline{\footnotesize W.N.G. HITCHON}
\vspace*{0.015truein}
\centerline{\footnotesize\it Department of Electrical Engineering, University 
of Wisconsin-Madison }
\baselineskip=10pt
\centerline{\footnotesize\it Madison, WI 53706, U.S.A.}
\vspace*{0.225truein}

\vspace*{0.21truein}
\begin{abstract}
Correlation effects of an electron gas in an external 
potential are derived using
an Effective Action 
 functional method. \ Corrections beyond the random phase approximation (RPA)
are naturally incorporated by this method. \ The Effective Action functional
 is made 
to depend explicitly on two-point correlation functions. \ The calculation 
is carried out at imaginary time. \  For a homogeneous electron gas, 
we  calculate the effect of exchange 
on the ring diagrams
at zero temperature and
show how to include some of the ladder diagrams. \ Our results agree
well with known numerical calculations. \ We conclude by
showing that this method is in fact a variant
of  the time dependent density functional method and suggest that it is 
suitable to be applied to the study of
correlation effects in the non-homogeneous case. 
\end{abstract}



\end{titlepage}

\section{Introduction }	
\vspace*{-0.5pt}
\noindent
Correlation effects  continue to be a popular subject due to the
importance of these interactions in physical processes in general. Electron systems
form an important subset of these systems. The nonlinearity that appears in
this type of problem complicates the calculation of any physical
properties of the system that strongly depend on the correlation between the
individual particles. In this paper we are mainly 
concerned with a particular problem, that  of calculating the ground state equilibrium energy of a
large Fermi gas in an external potential.\ At zero temperature the 
random phase approximation (RPA) is a
well known method which addresses this problem. \ Here we present a way to
improve on it, using an effective action method, which includes higher order
interactions and which also  applies  at finite
temperature and
 is applicable to non-homogeneous Fermi systems. Currently,
 such systems are frequently treated using  the density functional
method (DFT). \cite{kohn}$ ,$\cite{sham}  \ In DFT, the density 
of the system
plays a central role and the energy is a functional of it. Both the 
method we present here and DFT 
are extensions of the Thomas-Fermi
method. In DFT the density is found by self-consistently 
 solving  a
one-particle Schrodinger equation. Correlation effects are taken into account
by using the homogeneous result locally. The kinetic energy term is taken to
be that of a non-interacting system having the same density as the real system
under consideration. In spite of all these approximations, density functional
methods have proved to give better values for binding energies than
Hartree-Fock in atoms, molecules and solids in general. The method
works best for nearly homogeneous systems; surface effects are handled
poorly by this method. Attempts to include a gradient of the density
in the energy functional gave mixed results. \ Similarly, systems with 
magnetic impurities do not give consistently good results 
when using DFT. \ The overall success of the method is
therefore  not well understood. \ Here  
we propose a functional approach  in the spirit of DFT, that might offer
advantages when treating inhomogeneity. \ The problem becomes harder
to manage numerically in real cases. \ The method involves solving 
integro-differential equations which we solve in the homogeneous 
case. \ In the non-homogeneous case we suggest a possible strategy 
for a solution.

\ Functional 
methods have given interesting
results in many different areas of physics. \cite{domincis}$,$
 \cite{vasilev}$,$
\cite{cornwall}$,$ \cite{mossB}$^{,}$\cite{weinberg} \ They allow 
better treatment of 
non-perturbative effects and give a coherent treatment of zero and 
non-zero temperature properties. \ A method such as the one 
presented here can be extended to study non-equilibrium 
properties of conducting ferromagnetic systems
. \cite{rebei2} \  DFT is itself 
a functional method; \ Argaman and Makov gave a
simple  introduction to DFT based on this 
language. \cite{makov}$,$ \cite{valiev} \ Recently Kotani 
\cite{kotani}$,$\cite{kotani0}
proposed the use of correlation expressions 
based on the RPA approximation in {\it ab 
initio} calculations. \ His results were promising and consistent especially 
for magnetic elements. \ The method proved to be capable of giving 
better results than DFT in some cases. \ However, it was concluded that 
correlation expressions beyond the RPA are needed to give consistently better
answers than DFT. \ In this paper, we show how to get energy expressions
which are  
more accurate than those obtained 
from RPA. \ Such expressions might be useful in numerical
calculations such as the one carried out by Kotani. \cite{kotani} 
We show below how to compute the free energy of the
electron system.  The system is assumed to be in an external static potential
$V(\vec{x})$. The result that we get is very general and could be applied to
many different systems. \  It will be seen that the method provides a very compact
expression for the energy that goes beyond the RPA method in a natural way. \ As 
an example, we
apply our result for the energy to the homogeneous case in three 
dimensional space. \ We  explicitly
calculate the effect of taking exchange into account in the ring diagrams. Our
results agree well with results obtained
 using quantum Monte Carlo methods. \ The 
treatment is done at finite temperature; \cite{rebei} \ only at the end 
do we take the zero-temperature limit. \ Such results can be very useful 
for mesoscopic systems where correlation effects are important.

The paper is laid out as follows: In section 2 we introduce
the thermodynamic potential $\Omega$ for a system subject to external sources.
$\Omega$ therefore will be a functional of these sources. The functional
$\Omega$ is the generator of connected Green's functions at finite
temperature. Usually, one point source functions are used. Here, we will
instead introduce two-point functions . The introduction 
of two-point external sources
enables us to take into account the higher order corrections to the stationary
phase approximation of the partition function in a simpler way. By including
merely two diagrams (expressed in terms of the variables which are conjugate
to the external sources) in the Effective Action, we are able to 
obtain in a very compact way the
contribution to the energy 
of all the second order diagrams (and beyond) of the Coulomb
interaction.
This is the crucial advantage of this version of the Effective Action. The
diagrams which are included here are two-particle irreducible, in contrast to
the one-particle irreducible diagrams that we get when only one-point
functions are used in the Effective Action. However, before bringing in
the two-point correlation functions, we
introduce a new field to replace the quartic Coulomb interaction by the well
known Stratanovich-Hubbard transformation, so that all the interactions become
local. This new field is simply the Hartree potential as in the RPA
method. \cite{negele} \ We explicitly
introduce a term that describes correlation of the Hartree field at two
locations. The sources can also be taken to be instantaneous so we can get a
time-translation invariant solution.

In section 3, we calculate the finite temperature 
Effective Action $\Gamma$. \ This action is obtained by a Legendre 
transformation from $\Omega$. Therefore, we
introduce new variables conjugate to the sources. \ At the tree level, the
$\Gamma$ functional is simply the classical 
action of the system and is related to the
free energy of the system. For homogeneous systems, $\Gamma$ is known as the
effective potential. In our case $\Gamma$ will be a functional of three
variables, the Hartree potential $\varphi(\vec{r})$, the correlation function
$C(x,y)$ of the Hartree potential, and the Green's function $\rho_{\alpha
\beta}(x,y)$ of the Fermi field. If the external sources are taken to be
instantaneous, the points $x = (\tau_{1},\vec{r_{1}}) $ and $y = (\tau
_{2},\vec{r_{2}})$, correspond to the same temperature, $\tau_{1}=\tau_{2}$,
and the function $\rho_{\alpha\beta}(x,y)$ simply becomes the density matrix
of the system. $\alpha$ and $\beta$ are spin indices. Even though it is
possible to express the free energy in terms of the density matrix, it makes
the manipulations more cumbersome. 

In section 4, we solve for $\varphi(x)$ and $C(x,y)$ in terms of $\rho
_{\alpha\beta}(x,y)$. We are able to do this  because in reality
$\rho_{\alpha\beta}(x,y)$ is the only independent function of 
the problem. In this case we
get an expression for the Effective Action $\Gamma$ in terms of the density
only. Therefore getting the expression for $\Gamma$ in terms of $\rho
_{\alpha\beta}(x,y)$ is the essential result of this work from which other
important calculations can be made.

In section 5, we treat the zero temperature case. Because 
of the finite temperature approach, we have a set of diagrams, called 
 `anomalous',
in the sense that they should not be present at zero temperature. \ The 
contribution of these 
 diagrams vanishes at zero temperature. \ We  go beyond RPA to
include exchange effects on the ring diagrams. We also show that some of the
ladder terms are included in our approximation.

In section 6, we continue the treatment started in the previous section and
calculate in detail 
the contribution of the ring diagrams if they include exchange.

In section 7, we reexamine the non-homogeneous problem. We show how this
method is related to the DFT formalism. In fact it is shown that a statement
like the Hohenberg-Kohn theorem is trivially realized in this formalism.
Similarly the question of v-representability can be given an affirmative
answer within perturbation theory.

Finally, the last section is the conclusion.


\section{The Thermodynamic Potential $\mathbf{\Omega}$}
\noindent
In this section, we obtain an expression for the thermodynamical 
functional $\Omega$ to
`order' $\hbar^{2}$. \cite{matsubara} \  This 
functional is the logarithm of the partition functional
$Z$ of the system in the presence of external sources. The $Z$ functional is
written as an integral over all possible allowable paths of the different
fields with a weighting factor that depends on the value of the action 
along  the
given path. \ The functional $\Omega$ is simply the 
finite temperature generating functional of
connected Green's functions of the fields involved.
However because the Coulomb interaction, quartic in the Fermion field, is 
difficult
to integrate, we introduce an auxiliary field by means of the well known
Hubbard-Stratanovich transformation so that we can avoid the quartic
interaction and its nonlocal behavior. In other words, the procedure consists
of transforming the problem into one in which the electrons interact locally
with a Hartree-type potential. In this treatment, exchange energy terms will
show up directly because of the anti-commutation of the Fermi fields, which is
built into the calculation as will be described below. The correlation terms
are due to the quantum fluctuations around the Hartree potential.

In the following we show how  terms to order $\hbar^{2}$
in $\Omega$ can be obtained;  we
are including effects of second order in the \emph{full} Hartree field of the
theory. \ Therefore the equation of motion 
satisfied by $\rho(x,y)$ includes corrections  beyond
the Hartree-Fock approximation and involves second order exchange effects.

 A non-relativistic interacting electron gas in an external potential
$V(\vec{x})$ has the following Hamiltonian in the second quantized form:%

\begin{eqnarray}
H_{0}  & =&\int d^{3}x\left[  -\frac{1}{2}\Psi_{\alpha}^{\dagger}(x)\nabla
^{2}\Psi_{\alpha}(x)+V(x)\Psi_{\alpha}^{\dagger}(x)\Psi_{\alpha}(x)\right]
\nonumber\\
& & +\frac{1}{2}\int d^{3}xd^{3}y\frac{1}{|\vec{x}-\vec{y}|}\Psi_{\alpha
}^{\dagger}(x)\Psi_{\nu}^{\dagger}(y)\Psi_{\nu}(y)\Psi_{\alpha}(x).
\end{eqnarray}
Here, we use units such that $\hbar=m=e=k_{B}=1$ and $\beta
\equiv\frac{1}{T}$. $\Psi(x)$ is a two-component temperature-dependent
electron field Heisenberg operator. $\alpha$ and $\nu$ are 
spin indices, i.e., $\alpha=1$ for spin
up and $\alpha=-1$ for spin down. Summation is implicit for repeated indices.
In the following $\vec{x}$ will always mean a 3-D space vector. The system is
constrained by the condition%

\begin{equation}
\int d^{3}x\; \Psi_{\alpha}^{\dagger}(x)\Psi_{\alpha}(x)=N
\end{equation}
where N is the electron number operator which is constant. Therefore the
density operator is simply $\Psi_{\alpha}^{\dagger}(x)\Psi_{\alpha}(x).$
Because of this constraint, we prefer to work instead with the following Hamiltonian%

\begin{equation}
H=H_{0}-\mu N
\end{equation}
where $\mu$ is a Lagrangian multiplier. The electron operator $\Psi(x)$ can be
expanded in terms of a complete orthonormal set of one-particle functions
$\varphi_{k}(x),$ so we write
\begin{eqnarray}
\Psi(x)  & =&\sum_{k}a_{k}\varphi_{k}(x),\nonumber\\
\Psi^{\dagger}(x)  & =&\sum_{k}a_{k}^{\dagger}\varphi_{k}^{\ast}(x).
\end{eqnarray}
$a_{k}$ and $a_{k}^{\dagger}$ are annihilation and creation operators. The
subscript $k$ includes momentum and spin. The ground state and the excited
states can be represented as Slater determinants formed by the wave functions
$\varphi_{k}(x)$. In our case, it will be advantageous to take these functions
to be self-consistent Hartree eigenfunctions. In the homogeneous case they
reduce to plane waves. Since we are going to use a path integral formulation,
we give the Lagrangian associated with the Hamiltonian H,

\begin{eqnarray}
L  & = & \int\! d^{3}x\left[  i\Psi^{\dagger}(x) \frac{\partial
}{\partial t}\Psi(x)+\frac{1}{2}\Psi(x)\nabla^{2}\Psi^{\dagger}(x)-V(\vec
{x})\Psi^{\dagger}(x)\Psi(x)+\mu\Psi^{\dagger}(x)\Psi(x)\right] \nonumber\\
& & -\frac{1}{2}\int d^{3}xd^{3}y\frac{1}{|\vec{x}-\vec{y}|}\Psi^{\dagger
}(x)\Psi^{\dagger}(y)\Psi(y)\Psi(x) .
\end{eqnarray}

 For a system at thermal equilibrium, we have to replace the time
component t in the Hamiltonian by $-i\tau$. The partition function $Z$ is as
usual defined as a sum over all possible states $\Phi$. In the absence of
external sources, we have

\begin{eqnarray}
Z[0]  & = &\sum_{\Phi} \left\langle {\Phi} \, \vert\, e^{-\beta H}\, \vert\,
{\Phi}\right\rangle \nonumber\\
& = & N\int D\Psi D{\Psi}^{\dagger} \exp\left( {-S[\Psi,\Psi^{\dagger}]} \right)
\end{eqnarray}
where N is a normalization constant and the integration measure of the Fermi
fields is only defined for fields that satisfy \cite{martin}:
\begin{equation}
\Psi(0,x) = - \Psi(\beta,x) .
\end{equation}
$S[\Psi,\Psi^{\dagger}]$ is the action of the system,

\begin{eqnarray}
S[\Psi,\Psi^{\dagger}]  & = & \int_{0}^{\beta}d\tau\,L\nonumber\\
& =& \int_{0}^{\beta}d\tau\int d^{3}x\;\left \{ \Psi^{\dagger}(x)\left [ \frac
{-\partial}{\partial\tau}+\frac{1}{2}\nabla^{2}+\mu-V(x)\right ]
\Psi(x) \right \}
\nonumber\\
& & -\frac{1}{2}\int_{0}^{\beta}d\tau\int d^{3}xd^{3}y\;\frac{1}{|\vec{x}%
-\vec{y}|}\Psi^{\dagger}(x)\Psi^{\dagger}(y)\Psi(y)\Psi(x).
\end{eqnarray}\\
 The functional integral for the partition function becomes

\begin{eqnarray}
Z[0]  & =& N\int D\Psi D\Psi^{\dagger}\exp \left \{ -\int d\tau_{x}
d^{3}x \left [ \Psi^{\dagger}(x)\left(  \frac{\partial}{\partial\tau}-\frac
{1}{2}\nabla^{2}-\mu+V(\vec{x})\right)  \Psi(x) \right . \right . \nonumber\\
& &\mbox{} \left . \left . + \frac{1}{2}
\int_{0}^{\beta}d\tau\;d\tau_{y}\int d^{3}y
\frac{\Psi^{\dagger}(x)\Psi(x)\Psi^{\dagger}(y)
\Psi(y)}{|\vec{x}-\vec{y}|}\delta(\tau-\tau_{y}) \right ] \right \}.\label{e9}
\end{eqnarray}
In writing the last term we made use of the fact that the Fermi fields satisfy
a Grassmann algebra. An infinite self-energy term has been dropped from the
above expression. Such a term cancels at the end and has no effect on the
evaluation of the eigenfunctions or related physical quantities. \ For 
notational convenience we set

\begin{equation}
G^{-1}(x,y)=\left(  \frac{\partial}{\partial\tau_{x}}-\frac{1}{2}\nabla
^{2}-\mu+V(\vec{x})\right)  \delta(x-y)
\end{equation}
and

\begin{equation}
A(x,y)=\frac{\delta(\tau_{x}-\tau_{y})}{|\,\vec{x}-\vec{y}\,|}.
\end{equation}

Before proceeding further, we have to get rid of the quartic term in
the Lagrangian. This is done by introducing an auxiliary boson field 
 $ \varphi \left( x \right)$.\
We write%

\begin{eqnarray}
Z[0]  & =& N^{\prime}\int D\Psi D\Psi^{\dagger}D\varphi\;\exp \left \{-\int
d^{4}xd^{4}y\;\Psi_{\alpha}^{\dagger}(x)G^{-1}(x,y)
\Psi_{\alpha}(y) \right .\nonumber\\
& & \left . -\frac{1}{2}\int d^{4}xd^{4}y\;\varphi(x)A^{-1}(x,y)\varphi(y)+\int
d^{4}x\;\varphi(x)\Psi_{\alpha}^{\dagger}(x)\Psi_{\alpha}(x) \right \}.
\end{eqnarray}
 The fourth component is the time component integrated over the proper
range. It is easy to see that by using the following formula,%

\begin{equation}
\int dx\;e^{-\frac{1}{2}xQx+bx}=(\det Q)^{-\frac{1}{2}}\;e^{\frac{1}{2}%
bQ^{-1}b}%
\end{equation}
 and integrating over $\varphi$ we get back the original expression
for $Z\lbrack0 \rbrack$. The prefactor $N^{\prime}$ is a new normalization
constant. The operator $A(x,y)$ is clearly invertible,%

\begin{equation}
A^{-1}(x,y)=-\frac{1}{4\pi}\delta(x-y)\nabla^{2}.
\end{equation}
 The non-local character of the Coulomb interaction has been removed
by introducing the new bosonic field $\varphi(x).$ It can be shown that
$\varphi(x)$ is the Hartree potential by using the new equations of motion
derived from the new action of the problem.
Now  we  couple the fields to local
and non-local sources $J(x),Q(x,y)$ and $B(x,y)$. We first 
introduce the
functional $\Omega$, a generator of connected Green's functions,  through 
the normalized partition functional which is now a
functional of the external sources. It is given by:%

\begin{equation}
Z[J,B,Q] =\exp\left \{ -\beta\;\Omega[J,B,Q] \right \}
\end{equation}
such that

\begin{eqnarray}
\lefteqn{\exp\left(-\Omega\lbrack J,B,Q \rbrack\right)\int D\Psi D\Psi^{\dagger}D\varphi
\;\exp\left(  -S[\Psi,\Psi^{\dagger},\varphi]\right)   = } \nonumber\\
& & \int D\Psi D\Psi^{\dagger}D\varphi\exp \left \{ -S[\Psi,\Psi^{\dagger}
,\varphi]+\int d^{4}xd^{4}y\Psi^{\dagger}(x)Q(x,y)\Psi(y) \right . \nonumber \\
& &\mbox{} + \left . \frac{1}{2}\int d^{4}xd^{4}y\varphi(x)B(x,y)\varphi(y) 
+\int d^{4}x\varphi(x)J(x) \right \}
\end{eqnarray}
 where we have set $\beta = 1$.  
$S[\Psi,\Psi^{\dagger},\varphi]$ is simply the action of the
transformed problem,%

\begin{eqnarray}
S[\Psi,\Psi^{\dagger},\varphi]  & = & \int d^{4}xd^{4}y\;\Psi^{\dagger}%
(x)G^{-1}(x,y)\Psi(y)+\frac{1}{2}\varphi(x)A^{-1}(x,y)\varphi(y)\nonumber\\
& &-\int d^{4}x\varphi(x)\Psi^{\dagger}(x)\Psi(x) .
\end{eqnarray}

Now we define three new variables:%

\begin{equation}
\frac{\delta\Omega[J,B,Q]}{\delta J(x)}\vert_{J=B=Q=0}=\left\langle
\varphi(x)\right\rangle \equiv\varphi_{c}(x)
\end{equation}%

\begin{equation}
\frac{\delta\Omega\lbrack J,B,Q]}{\delta B(x,y)}|_{J=B=Q=0} =\frac{1}
{2}\left\langle \varphi(x)\varphi(y) \right\rangle \equiv\frac{1}{2}
\left [ \varphi_{c}(x)\varphi_{c}(y)+C(x,y) \right ]
\end{equation}%

\begin{equation}
\frac{\delta\Omega\lbrack J,B,Q]}{\delta Q(x,y)}|_{J=B=Q=0}=\left\langle
\Psi_{\alpha}^{\dagger}(x)\Psi_{\beta}(y)\right\rangle \equiv\rho_{\alpha
\beta}(x,y).
\end{equation}
 $\varphi_{c}(x)$ is the expectation value of the field $\varphi(x)$
in the ground state. $C(x,y)$ is a correlation function of the field
$\varphi(x).$ $\rho_{\alpha\beta}(x,y)$ is the Green function of the Fermi
field. $\rho_{\alpha\beta}(x,x)$ is therefore the density of the system. Here
and below, the ``time'' ordering operator is not written explicitly. Therefore
terms like $\rho(\vec{x},\vec{y})$ are defined by setting $\tau_{x}-\tau
_{y}=0^{+}.$ Note that $C(x,y)$ measures the departure from quasi-independence
due to the correlation between the values of the potential at two different
locations. The expectation value of the Fermi field $\Psi(x)$ is zero. 

In the
rest of this section, we  obtain an explicit expression for $\Omega\lbrack J,B,Q]$ to
`order' $\hbar^{2}$. \ In the next section   
we solve for $\varphi_{c}(x),C(x,y)$ and $\rho(x,y)$
in terms of $J(x),B(x,y)$ and $Q(x,y)$ to get an expression for the Effective
Action $\Gamma$ which is a functional of the new variables 
only.\ First we determine $\Omega$. \ We
expand the exponent in the above integral around $\Psi=\Psi^{\dagger}=0$ and
$\varphi=\varphi_{0}$ where $\varphi_{0}$ is the configuration of $\varphi$
that extremizes the action S. Therefore, we have%

\begin{equation}
\left(  A^{-1} + B\right)  \varphi_{0}=-J .
\end{equation}
 It is understood from the above that there is an integration over
space and time on the L.H.S. of this expression. We choose from now on not to
write integrals explicitly unless there might be some confusion. Now we expand
around $\varphi_{0}$, so we write

\begin{equation}
\varphi_{(old)}=\varphi_{(new)}+\varphi_{0} .
\end{equation}
\newline Then assuming that the main contribution to the integral comes from
the saddle point, we get the following 
expression for the partition functional:

\begin{eqnarray}
\exp\{-\beta\Omega\}&=&\exp(-\beta\;S[\varphi_{0}])\;\frac{\det[\widetilde
{G}^{-1}+Q]\;\det[A^{-1}+B]^{-\frac{1}{2}}}{\det[G^{-1}]\;\det[A^{-1}
]^{-\frac{1}{2}}}\,\Sigma^{-1} \nonumber  \\
&& \mbox{} \times\,\exp \left [ \int D\Psi D\Psi^{\dagger}D\varphi\;\;\Xi\; 
\left \{\frac{1}{2}e^{2}(\varphi\Psi^{\dagger}\Psi)^{2}\,\;+\ldots \right \}
\, \right ] \label{e23}
\end{eqnarray}
where we have defined $\Xi$ and $\Sigma$ to be
\begin{eqnarray}
\Xi & =&\exp \left [-\Psi^{\dagger}(\widetilde{G}^{-1}+Q)\Psi-\frac{1}%
{2}\varphi(A^{-1}+B)\varphi \right ] \\
\Sigma & =&\int D\Psi D\Psi^{\dagger}D\varphi\;\exp \left [-\Psi^{\dagger
}(G^{-1}+Q)\Psi-\frac{1}{2}\varphi(A^{-1}+B)\varphi \right ]
\end{eqnarray}
 In the above, we have used the fact that%

\begin{equation}
\int da\;da^{\dagger}\;e^{-a^{\dagger}Ma}=det\;M
\end{equation}
for Grassmann numbers a and a$^{\dagger}$, and%

\begin{equation}
\int dx\; e^{-\frac{1}{2}xMx}={\left(  det M \right) }^{-\frac{1}{2}}%
\end{equation}
for c-numbers x. The argument of the first term on the right is%

\begin{equation}
S[\varphi_{0}]= \frac{1}{2}\varphi_{0}A^{-1}\varphi_{0} + \frac{1}{2}
\varphi_{0}B\varphi_{0} + J\varphi_{0}%
\end{equation}
and%

\begin{equation}
\widetilde{G}^{-1}=G^{-1}-\varphi_{0} .
\end{equation}
\smallskip Now using the fact that%

\begin{equation}
\det A=e^{Tr\ln A},
\end{equation}
we can get an explicit expression for $\Omega[J,B,Q]$:%

\begin{eqnarray}
\Omega\lbrack J,B,Q]  & = & S(\varphi_{0})-\;Tr\ln[1+G\,\varphi_{0}+GQ]+\frac
{1}{2}\;Tr\ln[1+A\,B]\nonumber\\
& & -\frac{1}{2}\!\int\!D\Psi D\Psi^{\dagger}D\varphi\;(\varphi\Psi^{\dagger
}\Psi)^{2}\;\frac{\Xi}{\Sigma}+\ldots  \label{e31}
\end{eqnarray}
 After finding $\Omega$, we now calculate the Effective Action
$\Gamma$. We have to solve for the external sources in terms of the physical
variables $\rho$, $C$ and $\varphi_{c}$. This we do in the next section where
we find the Effective Action at finite temperature. However, this treatment
applies to zero temperatures too. In both cases  we get 
an expansion in $\hbar$
. \ However $\hbar$  is only used in bookkeeping 
 the diagrams included in the
expansion of $\Omega$.

\section{ The Effective Action $\mathbf{\Gamma}$ }
\noindent
The Effective Action is a functional of $\varphi_{c}(x)$, $C(x,y)$, and
$\rho(x,y)$, which is obtained by a triple Legendre transformation from
$\Omega\lbrack J,B,Q]$%

\begin{eqnarray}
\Gamma\lbrack\varphi_{c},C,\rho]  & =&\Omega\lbrack J,B,Q]-\int d^{4}%
x\;\varphi_{c}(x)J(x)-\frac{1}{2}\int d[xy]\;\varphi_{c}%
(x)B(x,y)\varphi_{c}(y)\nonumber\\
& & -\frac{1}{2}\int d[xy]\;C(x,y)B(x,y)-\int d[xy]\rho(x,y)Q(x,y)\; ,
\end{eqnarray}
where we have written $d$ $[xyz...]$ for $d^{4}xd^{4}yd^{4}z...$ to simplify
the notation. It is understood that the integration over time is carried out 
for
$\tau$ in the range 
$[0,\beta]$ for the nonzero temperature case and over all time
for the zero temperature case. 

It is easily verified that :%

\begin{equation}
\frac{\delta\Gamma[\varphi_{c},C,\rho]}{\delta\varphi_{c}(x)} =-J(x)-\int
d^{4}y\; B(x,y)\varphi_{c}(y) \label{e33}
\end{equation}%

\begin{equation}
\frac{\delta\Gamma\lbrack\varphi_{c},C,\rho]}{\delta C(x,y)}=-\frac{1}%
{2}B(x,y),\label{e34}
\end{equation}
and
\begin{equation}
\frac{\delta\Gamma\lbrack\varphi_{c},C,\rho]}{\delta\rho(x,y)}=-Q(x,y).\label{e35}
\end{equation}
When we turn off the external sources, the above equations give the
values of $\varphi_{c},C$ and $\rho$ that minimize the Effective Action.
$\varphi_{c}(x)$ and $C(x,y)$ are really dependent variables since they depend
on the auxiliary field $\varphi(x)$, so we should be able in principle to
express them in terms of $\rho(x,y)$ which gives us the density.

Now we have to express $J,B$ and $Q$ in terms of $\varphi_{c}$, $C$ and $\rho
$. First we note that when $\hbar=0,$ we have
\begin{equation}
\frac{\delta\Omega}{\delta J}=\varphi_{0} .
\end{equation}

Hence, we can write%

\begin{equation}
\varphi_{0}=\varphi_{c}+\widetilde{\varphi}%
\end{equation}
\medskip where $\widetilde{\varphi}$ is of order $\hbar$. Similarly, we write

\begin{equation}
S(\varphi_{0})=S(\varphi_{c})+ \hbar S_{1} .
\end{equation}
\medskip Now we seek an expression for $\Gamma$ in the form%

\begin{equation}
\Gamma=\Gamma_{0}+ \hbar \Gamma_{1} +\hbar^{2} \Gamma_{2} .
\end{equation}
\medskip It then follows that%

\begin{equation}
\Gamma_{0}=\frac{1}{2}\varphi_{c}A^{-1}\varphi_{c}.
\end{equation}
\medskip Now, we find approximate expressions for $B$ and $Q$ in terms of
$\varphi_{c},C$ and $\rho$. Using Eq.(\ref{e31}) and Eq.(\ref{e35}), we get 
the following relation:

\begin{equation}
\rho^{-1}G=1+GQ-eG\varphi_{c}+O(\hbar).
\end{equation}
\medskip The expression for $B$ is more involved. Again using Eq.(\ref{e35}) 
and
treating $\varphi_{0}$ as a functional of $J$ and $B$, we get:%

\begin{equation}
B\,C\,=\,(1-2e\,\rho\,\varphi_{c})\left(  1-2\frac{\varphi_{c}\widetilde
{\varphi}}{C}\right)  ^{-1}\,-A^{-1}\,C\,.
\end{equation}
\medskip Inserting back all these expressions in $\Gamma$, we get for
$\Gamma_{1}$

\begin{equation}
\Gamma_{1}=-Tr\;\ln\;(\rho^{-1}G)+\frac{1}{2}Tr\;\ln(C^{-1}A)+\frac{1}%
{2}Tr\;(A^{-1}C)-Tr\;(\rho\widetilde{G}^{-1}).
\end{equation}

\begin{figure}[htbp]
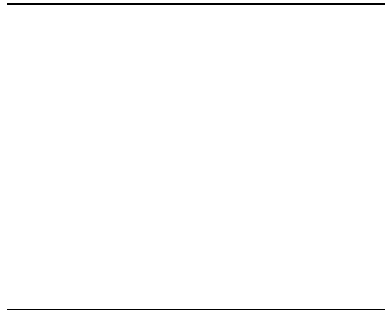

\vspace*{13pt}
\centerline{\vbox{\hrule width 5cm height 0.001pt}}
\vspace*{1.4truein}
\centerline{\vbox{\hrule width 5cm height 0.001pt}}
\vspace*{13pt}
\caption{Graphs a, b, c and d
are part of the expansion terms in $\Gamma$. Graph e is however reducible and
does not show up in the expansion. The solid line represents the propagator
$\rho(x,y)$, the dashed line represents the propagator $C(x,y)$.}%
\label{expansion}%
\end{figure}

 The terms of order $\hbar^{2}$ are also
straightforward but more involved. The steps 
are similar to those in the case of the Effective
Action with one-point sources only.  \cite{jackiw} \ Actually, 
it can be shown
that the next terms in $\Gamma_{2}$ are the sum of  two-particle irreducible
diagrams of the theory. \cite{vasilev}$, $\cite{cornwall}\ Here, we 
sum only the
first two diagrams in this series expansion. Fig.~\ref{expansion} shows the
first four diagrams in this expansion.\ The first two diagrams enable us to
include first and second order exchange effects. Diagram e is not part of
$\Gamma_{2}$ since it is two-particle reducible. However it is one-particle
irreducible and it is part of the usual Effective Action, \cite{negele}

\begin{eqnarray}
\Gamma_{2}  & = & -\frac{1}{2}e^{2}\int d[xy]\;\rho(x,y)\rho
(y,x)C(x,y)\nonumber\\
& & -\frac{1}{4}e^{4}\int d[xyuv]\;\rho(x,y)\rho(y,u)\rho(u,v)\rho
(v,x)C(x,u)C(y,v).
\end{eqnarray}

\noindent Hence the reducible graphs do not appear in the expansion. It is the
term $\Sigma$ that appeared in Eq.(\ref{e23}) that is responsible for excluding such
a graph from the expansion. $\Gamma_{2}$ and higher order terms represent
vacuum graphs where the {\it full }propagators 
are $\rho(x,y)$ and $C(x,y)$ of the Fermi
field and the boson field, respectively. Therefore to order $\hbar^{2}$, the
Effective Action has the following expression%

\begin{eqnarray}
\Gamma\lbrack\varphi_{c},C,\rho]  & =& -S(\varphi_{c})+\frac{1}{2}Tr\;\ln
AC^{-1}+A^{-1}C\\
& & -Tr\;\ln\rho^{-1}G+\widetilde{G}^{-1}(\varphi_{c})\rho+\Gamma_{2}%
[C,\rho],\nonumber
\end{eqnarray}
\newline where

\begin{equation}
\widetilde{G}^{-1}(\varphi_{c})=\left(  \frac{\partial}{\partial\tau}-\frac
{1}{2}\nabla^{2}-\mu+V(x)-\varphi_{c}(x)\right)  .
\end{equation}
\newline The term $\mu\int d^{3}x$ $\rho(x,x)$ cancels the term $\mu N$ in the
Hamiltonian H, and therefore it can be handled without difficulty. However, we
have to deal with the $\mu$ that appears in the term $Tr\,\ln\rho^{-1}\,G$.
This can be removed by a convenient choice of the path of integration in the
complex $\omega$-plane. The case $\mu=0$ occurs when there are no charges
present. We are dealing with negatively charged electrons bound by a static
potential $V(x)$. Therefore bound states appear, with negative energies
bounded from below for any physically sensible $V(x)$. In Fig.~\ref{path} we
show how by going from the $C_{1}$ path, which corresponds to nonzero $\mu$,
to the $C_{0}$ path we pick up contributions from the bound states between the
two paths, since energies of bound states are poles of the propagator $G^{-1}$
in the complex $\omega$ plane. Note that the eigenvalues $\omega_{k}$ of these
states do not take into account correlations due the Coulomb field. Because of
the logarithmic operator, there is a cut along the real positive axis. By
going back to real time and then Fourier transforming the 
time dependence, we get
\begin{eqnarray}
 \int_{C_{1}}dt\;Tr\ln\left(  -i\frac{\partial}{\partial t}-\frac{1}{2}%
\nabla^{2}-\mu+V(x)\right) \nonumber\\
 =\int dt\int_{C_{0}}\frac{d\omega}{2\pi i}\;tr\;\ln\left(  -\omega-\frac
{1}{2}\nabla^{2}+V\right)  +\sum_{k}\omega_{k}\label{e47}
\end{eqnarray}
\newline where $tr$ applies only to spatial variables. The $\omega_{k}$'s are
the eigenvalues of the  equation

\begin{equation}
\left(  -\frac{1}{2}\nabla^{2}+V(x)\right)  \varphi_{k}(x)=\omega_{k}%
\varphi_{k}(x).
\end{equation}
\newline The single particle functions $\varphi_{k}(x)$ of the potential
$V(x)$ are not supposed to be used alone as the starting point for any
numerical calculations since they do not take account of the repulsion between
electrons. However, it is clearly 
advantageous to separate the effect of the external
potential $V(x)$. This separation also appears in other treatments such as DFT.

The ground state energy E$_{g}$ is given in terms of the Effective Action per
unit time. Using the above results, and the fact that the time independent
values of $\rho(x,y),C(x,y)$ and $\varphi_{c}(x)$ are defined by setting
$\tau_{x}=\tau_{y}=0$, the expression for $E_{g}$ is%

\begin{eqnarray}
E_{g}\int dt  & = &\frac{1}{2}\int d[xy]\;\varphi_{c}(x)A^{-1}(x,y)\varphi
_{c}(y)\;+\frac{1}{2}Tr\;\ln C\;A^{-1}\nonumber\\
& & -\frac{1}{2}\int d[xy]\;A^{-1}(x,y)C(y,x)\;-\sum_{\mu<\omega_{k}<0}%
\omega_{k}\int dt\nonumber\\
& & -Tr\;\ln\rho G^{-1}+\int d^{4}x\left(  \frac{\partial}{\partial\tau_{x}%
}-\frac{1}{2}\nabla^{2}+V(x)-e\;\varphi_{c}(x)\right)  \rho(x,y)|_{\tau
_{x}=\tau_{y}}^{x=y}\nonumber\\
& & +\frac{1}{4}e^{4}\int d[xyuv]\;\rho(x,y)\rho(y,u)\rho(u,v)\rho
(v,x)C(x,u)C(y,v)\nonumber\\
& & +\frac{1}{2}e^{2}\int d[xy]\;C(x,y)\rho(x,y)\rho(y,x)\label{e49}
\end{eqnarray}
where we have dropped terms of order higher than $e^{4}$ in $\Gamma_{2}$ for
simplicity. This is the full expression for the ground state energy of the
system. Here $G^{-1}$ has $\mu$ set to zero. To be able to use this expression
for $E_{g}$, more approximations 
are required.\  Since $\varphi_{c}$ is a dependent field, in
principle we should be able to solve for $\varphi_{c}$ and $C$ in terms of
$\rho$ only. This we do next, but only approximately to keep the calculations manageable.

\begin{figure}[ptb]
\vspace*{13pt}
\centerline{\vbox{\hrule width 5cm height 0.001pt}}
\vspace*{1.4truein}
\centerline{\vbox{\hrule width 5cm height 0.001pt}}
\vspace*{13pt}
\caption{Path of integration used to obtain Eq.(\ref{e47}).}
\label{path}
\end{figure}

\newpage

\section{The Effective Action as a Functional of $\mathbf{\rho(x,y)}$}
\noindent
In this section, we find an expression for $\Gamma$ solely in terms of the
Green's function $\rho(x,y).$ This is done by finding the expressions for
$\varphi_{c}(x)$ and $C(x,y)$, that minimize $\Gamma$, in terms of the density
$\rho(x,y).$ For $\varphi_{c}(x)$ we get the following expression:%

\begin{equation}
\varphi_{c}(x)=-e\,\int d^{3}y\,\frac{\rho(y,y)}{|\vec{x}-\vec{y}|}.\label{e50}
\end{equation}

\noindent Similarly, minimizing $\Gamma$ with respect to $\rho(x,y)$, we get%

\begin{eqnarray}
\delta(x,z)&=&\left[  \frac{\partial}{\partial\tau_{x}}-\frac{1}{2}\nabla
^{2}+V(x)-e\;\varphi_{c}(x)\right]  \rho(x,z)+e^{2}\int d^{4}y\,\rho
(x,y)C(x,y)\rho(y,z)\nonumber \\
&& -e^{4}\int d[yuv]\rho(v,u)\rho(u,x)\rho(y,v)\rho(z,y)C(u,y)C(v,x)\: .\label{e52}
\end{eqnarray}
\newline Finally, minimizing $\Gamma$ with respect to C, we get%

\begin{eqnarray}
\delta(x,z)&=&\int d^{4}y \left [ A^{-1}(x,y)-e^{2}\rho(x,y)\rho(y,x)\right ]
C(y,z)\nonumber \\
&& -e^{4}\int d[yvu]C(u,v)C(y,z)\rho(x,u)\rho(u,y)\rho(y,v)\rho(v,x)\: .\label{e53}
\end{eqnarray}
To be able to solve for $C(x,y)$ we have to linearize the theory. In the
following, we keep terms only to `order' $e^{4}$. \ We find that the
correlation $C(x,y)$ is given by%

\begin{eqnarray}
C(x,z)&=&A(x,z)+e^{2}\int d[x_{1}x_{2}]\;A(x,x_{1})D(x_{1},x_{2})
A(x_{2},z)\nonumber\\
&& +e^{4}\int d[x_{1}x_{2}x_{1}^{\prime}x_{2}^{\prime}]\;A(x,x_{1}^{\prime
})D(x_{1}^{\prime},x_{2}^{\prime})A(x_{2}^{\prime},x_{1})D(x_{1},x_{2}%
)A(x_{2},z)\nonumber \\
&& +e^{4}\! \! \int \! d[x_{1}x_{2}x_{1}^{\prime}x_{2}^{\prime}]\;\rho(x_{1},x_{1}%
^{\prime})\rho(x_{1}^{\prime},x_{2})\rho(x_{2},x_{2}^{\prime})\rho
(x_{2}^{\prime},x_{1})A(x_{2},x_{1})A(x_{2}^{\prime},z)A(x,x_{1}^{\prime
})\label{e54}
\end{eqnarray}
 where we have set

\begin{equation}
D(x_{1},x_{2})=\rho(x_{1},x_{2})\rho(x_{2},x_{1})\: .
\end{equation}
 The above equation for $C(x,z)$ can be represented diagrammatically
as shown in Fig.~\ref{Cex}. The first term is the bare Coulomb potential.
The second and the third are the direct and exchange term with one bubble. So
far the Fermion propagators are the true propagators in this expansion. In the
following we will ignore the corrections with two bubbles and higher.

\begin{figure}[ptb]
\vspace*{13pt}
\centerline{\vbox{\hrule width 5cm height 0.001pt}}
\vspace*{1.4truein}
\centerline{\vbox{\hrule width 5cm height 0.001pt}}
\vspace*{13pt}
\caption{The expansion of the propagator
$C(x,z)$ to order $e^{4}$ .}%
\label{Cex}%
\end{figure}

\begin{figure}[ptb]
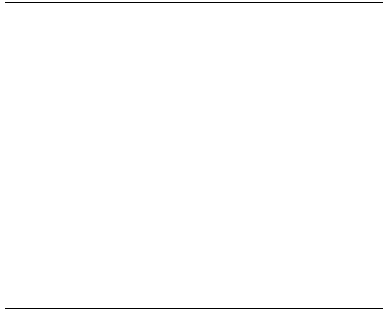

\vspace*{13pt}
\centerline{\vbox{\hrule width 5cm height 0.001pt}}
\vspace*{1.4truein}
\centerline{\vbox{\hrule width 5cm height 0.001pt}}
\vspace*{13pt}
\caption{Born approximation for
two-particle Green's function. }%
\label{ladder}%
\end{figure}

Clearly $C(x,y)$ is the full screened Coulomb potential if all
orders of $e^{2}$ are included. Similarly, to order $e^{4}$, the equation
satisfied by $\rho(x,z)$ is%

\begin{eqnarray}
\left [\frac{\partial}{\partial\tau_{x}}-\frac{1}{2}\nabla^{2}
+V(\vec{x})+e^{2}\int dy\frac{\rho(y,y)}{|\vec{x}-\vec{y}|} \right ]
\rho(x,z) = \delta(x,z) \nonumber \\
-e^{2}\int dy\,A(x,y)\rho(x,y)\rho(y,z)-e^{4}\int d[yx_{1}x_{2}]\;\rho
(x,y)\rho(y,z)A(x,x_{1})D(x_{1},x_{2})A(x_{2},y)\nonumber \\
+e^{4}\int d[yx_{1}x_{2}]\;\rho(x_{2},x_{1})\rho(x_{1},y)\rho(x,x_{2}%
)\rho(y,z)A(x_{1},x)A(x_{2},y)\: . \label{e56}
\end{eqnarray}
This is the  equation satisfied by the Fermi Green's functions
.\cite{negele}\ The term on the R.H.S. 
of order $e^{2}$ is an exchange term or
Fock term due to the statistics of the electrons. The last two terms of order
$e^{4}$ take into account collisions and are equivalent to the usual Born
approximation for two-particle Green's functions in 
scattering theory, Fig.~\ref{ladder}.

 The most likely way to solve these
integro-differential equations is by iteration. Using the above equations, we
can write an explicit expression for the energy in terms of the full
propagator $\rho(x,z)$ of the theory. However, such an expression suffers from
the problem of over-counting some of the states of the system. This is mainly
due to the fact that $e^{2}$ is not a good expansion parameter. Hence physical
arguments are used to discard some terms at this level of approximation and
instead include them at higher orders. 

\begin{figure}[ptb]
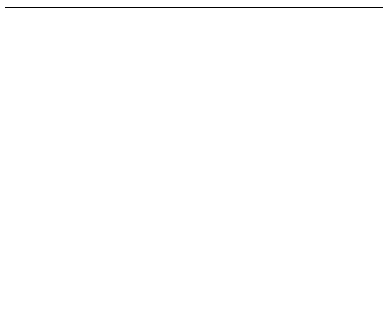

\vspace*{13pt}
\centerline{\vbox{\hrule width 5cm height 0.001pt}}
\vspace*{1.4truein}
\centerline{\vbox{\hrule width 5cm height 0.001pt}}
\vspace*{13pt}
\caption{Approximate
solution to the one-particle Green's function. The propagators on the 
right correspond to the free theory.}%
\label{rho-approx}%
\end{figure}

An expression for the electron
propagator in terms of the free propagator is given in Fig.~\ref{rho-approx}
. The corresponding analytical 
expression can be easily written following
usual rules \cite{negele}. Here we keep 
only the first three terms. The second
term vanishes in the homogeneous case, the case we are mainly 
interested in this
paper. We expect the expression for the energy, which has not been linearized,
to be a good one if we believe that a stationary phase approximation is
viable. Without including $\Gamma_{2}$, the expression obtained for the energy
is correct with $\rho(x,y)$ the Hartree propagator and $\varphi_{c}(x)$ the
Hartree potential. This indicates that a stationary phase approximation is
possible and that the higher order correction $\Gamma_{2}$ should be a small
perturbation to the Hartree solution. The linearization of the problem must
take account of this. Hence, we can justify the validity 
of the expansion of  $\Gamma$ in $\hbar$,

\begin{equation}
\Gamma=\Gamma_{0}+\hbar \Gamma_{1}+\hbar^{2}\Gamma_{2}+...
\end{equation}
by the result we get in the end.
 To second order in $\hbar$, Eq.(\ref{e54}) becomes
\begin{eqnarray}
\int dy A^{-1}(x,y)\;C(y,z)&=&\delta(x,z)+\hbar e^{2}\int\: dyD(x,y)A(y,z)
\nonumber \\
&& +\hbar e^{4}\int d[yuv]\: \rho(x,u)\rho(u,y)\rho(y,v)\rho
(v,x)A(u,v)A(y,z)\nonumber \\
&& -2\hbar^{2}e^{4}\int d[yuv]\rho(x,u)\rho(u,v)\rho(v,y)\rho(y,x)A(u,v)A(y,z)
\end{eqnarray}
where now $\rho(x,y)$ is the non-interacting electron propagator. We have used only a first
order approximation to the true propagator to find this equation for $C(x,y)$.
Using this equation, we get an expression for the energy to order $\hbar^{2}$ with
proper symmetry factors for the diagrams involved in the expansion. Setting
$\hbar=1$, we obtain for the zero temperature limit,%

\begin{eqnarray}
T\;E_{g}[\rho]& = &\frac{e^{2}}{2}\int dx\,dy\;\frac{\rho_{H}(x,x)\rho_{H}
(y,y)}{|\vec{x}-\vec{y}|}\,+\,e\int dx\,V(x)\rho_{H}(x,x) \nonumber \\
&& +\int dx\left(  -\frac{1}{2}\nabla^{2}\rho_{H}(x,y)\right)  |_{x=y}
\;-T\sum_{\omega_{k}>\mu}\omega_{k}\nonumber \\
&& -\frac{1}{4}\;e^{4}\!\int dx_{1}\,dx_{2}dx\,dy\;\rho_{H}(x_{1},x)\rho
_{H}(x,x_{2})\rho_{H}(x_{2},y)\rho_{H}(y,x_{1})A(x,y)A(x_{2},x_{1})\nonumber \\
&& +\frac{1}{2}\;Tr\,\ln \left [\delta(x,z)+e^{2}\int dyA(z,y)D_{H}
(x,y)+ \right . \nonumber \\
&& e^{4}\int dydx_{1}dx_{2}\;\rho_{H}(x_{1},x)\rho_{H}(x,x_{2})\rho_{H}
(x_{2},y)\rho_{H}(y,x_{1})A(x_{2},x_{1})A(y,z)\;- \nonumber \\
&& \left . 2e^{4}\int dydx_{1}dx_{2}\;\rho_{H}(x,x_{1})\rho_{H}(x_{1},x_{2})\rho
_{H}(x_{2},y)\rho_{H}(y,x)A(x_{2},x_{1})A(y,z)\;\right ] \nonumber \\
&& -Tr\,\ln \left [\;\delta(x,z)-e^{2}\!\int dy\;\rho_{H}(y,y)\rho
_{H}(x,z)A(x,y)-e^{2}\!\int dy\;A(x,y)\rho_{H}(x,y)\rho_{H}(y,z) \right .\nonumber \\
&& -e^{4}\int dydx_{1}dx_{2}\;\rho_{H}(x,y)\rho_{H}(y,z)\rho_{H}(x_{1},x_{2}%
)\rho_{H}(x_{2},x_{1})A(x_{1},x)A(x_{2},y) \nonumber \\
&& \left [+e^{4}\int dydx_{1}dx_{2}\;\rho_{H}(x_{2},x_{1})\rho_{H}(x_{1},y)\rho
_{H}(x,x_{2})\rho_{H}(y,z)A(x_{1},x)A(x_{2},y)\; \right ]\nonumber \\
&& -e^{2}\!\int dy\;\rho_{H}(y,y)\rho_{H}(x,z)A(y,x)|_{x=z}+e^{2}\int
dxdy\;A(x,y)\rho_{H}(x,y)\rho_{H}(y,z)\;|_{x=z} \nonumber \\
&& -e^{4}\!\int dxdydx_{1}dx_{2}\delta(x-z)\;A(x,x_{1})A(y,x_{2})\rho
_{H}(y,z)\rho_{H}(x_{2},x_{1})\rho_{H}(x_{1},y)\rho_{H}(x,x_{2})\nonumber \\
&& +e^{4}\int dxdydx_{1}dx_{2}\;\delta(x-z)\rho_{H}(x,x_{1})\rho_{H}(x_{1}%
,x_{2})\rho_{H}(x_{2},y)\rho_{H}(y,x)A(x_{2},x_{1})A(y,z) .\label{e59}
\end{eqnarray}
T is an interval of time. \ The trace, $Tr$, acts on x and z. So now we have obtained an
expression for the energy in terms of the Green's functions $\rho(x,z)$ only.
This is a major result of the work. As we mentioned above, we solve for
$\rho(x,y)$ to order $e^{4}$ by 
iteration of Eqs.(\ref{e52},\ref{e53}). The expression for
the energy can be solely written in terms of $\rho(x,x)$. However as we stated
earlier this is not advantageous and it is better to keep on 
working in terms of
$\rho(x,y)$.

 The classical Coulomb term and the interaction with the external
potential appear naturally in this approximation. They can be easily
separated from the full expression for the energy.
Another important point is
that a gradient of the density also appears naturally within the above
expressions. \ By expanding the first
logarithm, we immediately obtain the Hartree-Fock exchange term, i.e.,%

\begin{equation}
E_{exch}=\frac{1}{2}e^{2}\int dx\,dy\frac{\rho_{H}(x,y)\rho_{H}(y,x)}{|\vec
{x}-\vec{y}|}\delta(\tau_{x}-\tau_{y}).
\end{equation}
 The usual expression for the exchange energy follows trivially from
Eq.$\left(\ref{e59} \right)$ upon using the commutation relation of the Fermi
field. The infinity arising from this cancels the self-interaction term that
was dropped in Eq.(\ref{e9}). There is a 
similar exchange term that comes from the
second logarithmic term . This term is canceled by another similar term
outside the logarithm. The usual RPA term is contained in the first logarithm.
The new, extra term in the same logarithm provides among other things exchange
corrections to the ring diagrams. The higher order exchange diagrams are also
included in this approximation. We will say more on this when we treat the
homogeneous case in the following section.

\section{The Homogeneous Electron Gas at Zero Temperature}
\begin{figure}[ptb]
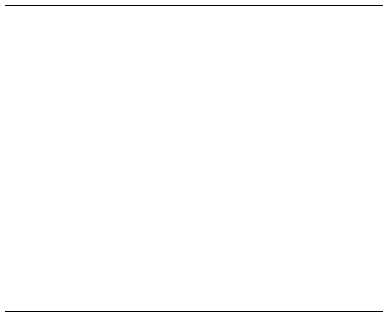

\vspace*{13pt}
\centerline{\vbox{\hrule width 5cm height 0.001pt}}
\vspace*{1.4truein}
\centerline{\vbox{\hrule width 5cm height 0.001pt}}
\vspace*{13pt}
\caption{A term of order
$e^{4}$ that is due to $\Sigma_{1}$. Both representations are equivalent. 
Initially two particles interact with each other with one of them 
going back to its
initial state while the other one interacts with a third particle before
both returning to their corresponding initial states.  }
\label{sigma1}%
\end{figure}

\begin{figure}[ptb]
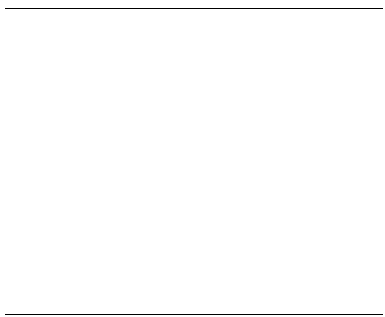

\vspace*{13pt}
\centerline{\vbox{\hrule width 5cm height 0.001pt}}
\vspace*{1.4truein}
\centerline{\vbox{\hrule width 5cm height 0.001pt}}
\vspace*{13pt}
\caption{ Some of the ladder
diagrams due to $\Sigma_{1}$.}%
\label{ladder3}%
\end{figure}

\noindent
In this section we apply the main result of section 4, the expression for the
energy Eq.(\ref{e59}), to the homogeneous 
case at zero temperature. \ The literature on this problem
of the calculation of the correlation energy is huge. \ The most complete
treatment, so far as we are aware, was given by Bishop and Luhrmann who used 
a linked cluster type of expansion 
 which was restricted to zero 
temperature effects. \cite{bishop} \ We assume that there is
a background of positive charge with density equal to the average density
of the electron gas, i.e., the system is neutral. \  Since the
system is homogeneous, the final expression for the energy will be given in
momentum space. The Green's function that will be used in the following is the
solution to the first iteration of the nonlinear equation satisfied by
$\rho(x,y)$. Since there is no external potential, the input Green's function
is that of a free electron gas. The energy expression is given in imaginary
time, hence we use the following expression for the free Green's function%

\begin{equation}
\rho_{0}(\omega,k)=\frac{1}{-i\omega+\frac{1}{2m}k^{2}-\mu} \: \: .
\end{equation}
\newline In this section and the next, we use this definition of 
the electron Green's function
which  differs by a factor of i from the one used previously. \ In terms of
$\rho_{0}(x,y)$ the ground state energy is given by the following (omitting
the subscript 0 for simplicity)


\begin{eqnarray}
-T\;E_{gs}&=&\int d^{4}x\,\left(  \frac{1}{2}i\nabla^{2}\rho(x,y)\right)
_{x=y}\nonumber \\
 &&+\frac{e^{4}}{4}\int dx\,dz\,\delta(x-z)\;{\mathcal{M}}(x,z)+\frac
{e^{2}}{2}\int dx\,dy\;\rho(x,y)\rho(y,x)A(x,y)\nonumber\\
&&-\frac{1}{2}Tr_{(x,z)} \ln\left[  \delta(x,z)-e^{2}%
{\mathcal{H}}(x,z)+ e^{4}\,{\mathcal{M}}(x,z)-2 e^{4}\Pi(x,z)\right]\nonumber\\
&& +Tr_{(x,z)}\;\ln\left[  \delta(x,z)-e^{2}{\mathcal{K}}(x,z)-e^{4}%
\;\Sigma(x,z)\right]\nonumber\\
&& +\int dx\,dz\,\delta(x-z)\;\left\{  \frac{e^{2}}{2}{\mathcal{H}}(x,z)-\frac
{e^{4}}{2}{\mathcal{M}}(x,z)+e^{4}\Sigma(x,z)+e^{4}\Pi(x,z)\right\}, \label{e62}
\end{eqnarray}
where
\begin{equation}
{\mathcal{M}}(x,z)=\int dy\,dx_{1}\,dx_{2}\,\rho(y,x_{2})\rho(x_{2}%
,x)\rho(x,x_{1})\rho(x_{1},y)A(x_{1},x_{2})A(y,z),
\end{equation}%
\begin{equation}
{\mathcal{H}}(x,z)=\int dy\,A(z,y)D(x,y),
\end{equation}%
\begin{equation}
\Pi(x,z)=\int dy\,dx_{1}\,dx_{2}\,\rho(y,x)\rho(x,x_{1})\rho(x_{1},x_{2}%
)\rho(x_{2},y)A(x_{1},x_{2})A(y,z),
\end{equation}%
\begin{equation}
{\mathcal{K}}(x,z)=\int dy\,A(x,y)\rho(x,y)\rho(y,z),
\end{equation}%
\begin{equation}
\Sigma(x,z)=\Sigma_{1}(x,z)-\Sigma_{2}(x,z),
\end{equation}%
\begin{equation}
\Sigma_{1}(x,z)=\int dy\,dx_{1}\,dx_{2}\,\rho(x,y)\rho(y,z)\rho(x_{1}%
,x_{2})\rho(x_{2},x_{1})A(x,x_{1})A(x_{2},y),
\end{equation}%
\begin{equation}
\Sigma_{2}(x,z)=\int dy\,dx_{1}\,dx_{2}\,\rho(x,x_{2})\rho(x_{2},x_{1}%
)\rho(x_{1},y)\rho(y,z)A(x_{1},x)A(x_{2},y).
\end{equation}
\newline The above equation, Eq.(\ref{e62}), is simply 
Eq.(\ref{e59}) taking into account of
the fact that $\rho(x,x)$ gets canceled by an equal and opposite charge.
Because of the translational invariance of the system, we can write a simple
expression for the correlation energy in the momentum representation,%

\begin{eqnarray}
E_{c}  & = & \frac{1}{4}e^{4}V\int\prod_{i=1}^{3}\frac{d^{3}p_{i}}{(2\pi)^{3}%
}\frac{d\omega_{i}}{2\pi}A(p_{1}-p_{2})A(p_{2}-p_{3})\rho(p_{1})\rho
(p_{2})\rho(p_{3})\rho(p_{1}-p_{2}+p_{3})\nonumber\\
&& +\frac{1}{2}V\int\frac{d^{3}p}{(2\pi)^{3}}\frac{d\omega}{2\pi} \left \{
\ln \left [ 1+e^{2}{\mathcal{H}}(p)+e^{4}{\mathcal{M}}(p)-2e^{4}\Pi
(p)\right ]-e^{2}{\mathcal{H}}(p)-e^{4}{\mathcal{M}}(p)+2e^{4}\Pi(p)\right \}
\nonumber\\
& & -V\int\frac{d^{3}p}{(2\pi)^{3}}\frac{d\omega}{2\pi}\left\{  \;\ln
\left [1-e^{2}{\mathcal{K}}(p)-e^{4}\Sigma(p) \right ] +e^{2}{\mathcal{K}}
(p)+e^{4}\Sigma(p)\;\right\}  \;.\label{eq:Ec}
\end{eqnarray}
\newline Since the ring diagrams are part of $E_{c}$, we can
test Eq.$\left( \ref{eq:Ec} \right)$   to see if our 
expansion gives the correct leading result. \ First we show
how this equation follows from Eq.(\ref{e62}). We start by letting%

\begin{equation}
{\mathcal{H}}(x,z)=-\int\,dy\: D(x,y)A(y,z)
\end{equation}
or, since the system is homogeneous, we can write instead%

\begin{equation}
{\mathcal{H}}(x-z)=\int dy\,D(x-y)A(y-z).
\end{equation}
\newline Next, we rewrite $D(x-y)$ and $A(x-z)$ in terms of their
corresponding Fourier transforms, i.e.,%

\begin{equation}
D(x-y)=\int\frac{dk}{(2\pi)^{4}}\;\exp[ik\cdot(x-y)]D(k)
\end{equation}
and
\begin{equation}
A(y-z)=\int\frac{dq}{(2\pi)^{4}}\;\exp[iq\cdot(y-z)]A(q).
\end{equation}
\newline Hence, by convolution we have,%

\begin{equation}
{\mathcal{H}}(p)\,=\,(A\,\star\,D)\;(p),
\end{equation}
\newline however since
\begin{equation}
D(p)=\int\frac{dq}{(2\pi)^{4}}\rho(p+q)\,\rho(q),
\end{equation}
\newline it follows then that,%

\begin{equation}
{\mathcal{H}}(p)\,=\, A(p)\,\int\frac{dq}{(2\pi)^{4}} \, \rho(p+q) \rho(q) .
\end{equation}
\newline After using the expressions for the propagators, we end up with the
following expression for ${\mathcal{H}}(p)$%

\begin{equation}
{\mathcal{H}}(p)=\frac{4\pi}{p^{2}}\!\int\!\frac{dq}{(2\pi)^{4}}\,\frac
{1}{-i\bar{\omega}+\frac{1}{2}\,{\vec{q}}^{2}-\mu} \; \frac{1}{-i(\bar{\omega
}+\omega)+\frac{1}{2}\,(\vec{p}+\vec{q})^{2}-\mu}%
\end{equation}
with $q=(\bar{\omega},\vec{q})$. Integrating first over $\bar{\omega
}$ in the complex plane, we find that%

\begin{equation}
{\mathcal{H}}(p)=-2e^{2}\frac{4\pi}{p^{2}}\!\!\int\!\!\frac{d\vec{q}}{(2\pi
)^{3}}\frac{\left(  \frac{1}{2}(\vec{p}+\vec{q})^{2}-\frac{1}{2}\vec{q}%
^{2}\right)  \left(  \Theta(\mu-\frac{1}{2}\vec{q}^{2})-\Theta(\mu-\frac{1}%
{2}(\vec{p}+\vec{q})^{2})\right)  }{\omega^{2}+\left(  \frac{1}{2}(\vec
{p}+\vec{q})^{2}-\frac{1}{2}\vec{q}^{2}\right)  ^{2}}\label{eq:AD}%
\end{equation}
 where we added a factor of $2$ to account for spin. This term
${\mathcal{H}}(p) $ will prove to be all that is needed to reproduce the known
RPA result. All other terms that appear in the energy expression are new
additions to the correlation. Before showing this explicitly , we give the
expressions for the remaining terms ${\mathcal{K}}(p)$,  ${\mathcal{M}}(p)$, 
$\Pi(p)$,  $\Sigma_{1}(p)$,  $\Sigma_{2}(p)$ and $\Sigma(p)$.

\begin{equation}
{\mathcal{K}}(p)=\rho(p)\int\frac{dq}{(2\pi)^{4}}\rho(p+q)A(q)
\end{equation}%
\begin{equation}
{\mathcal{M}}(p)=A(p)\int\frac{dk_{1}}{(2\pi)^{4}}\frac{dk_{2}}{(2\pi)^{4}}%
\rho(k_{1})\rho(k_{1}+p)\rho(k_{2})\rho(k_{2}-p)A(k_{1}-k_{2}+p),
\end{equation}%
\begin{equation}
\Pi(p)=A(p)\int\frac{dk_{1}}{(2\pi)^{4}}\frac{dq}{(2\pi)^{4}}\rho(k_{1}%
)\rho(k_{1}-p)\rho(k_{1}-p-q)A(q),
\end{equation}%
\begin{equation}
\Sigma_{1}(p)=\rho(p)\int\frac{d^{4}p_{1}}{(2\pi)^{4}}\frac{d^{4}p_{2}}%
{(2\pi)^{4}}\,\rho(p_{1})\rho(p_{2})\rho(p_{2}+p_{1}-p)A(p_{2}-p)A(p-p_{1}%
),
\end{equation}%
\begin{equation}
\Sigma_{2}(p)=\rho(p)\int\frac{d^{4}p_{1}}{(2\pi)^{4}}\frac{d^{4}p_{2}}%
{(2\pi)^{4}}\,A^{2}(p_{1})\rho(p-p_{1})\rho(p_{1}+p_{2})\rho(p_{2})
\end{equation}
and
\begin{equation}
\Sigma(p)=\rho(p)\int\frac{d^{4}k_{1}}{(2\pi)^{4}}\frac{d^{4}k_{2}}{(2\pi
)^{4}}\rho(k_{1})\rho(k_{2})\rho(k_{2}+k_{1}-p)\left[  A(k_{1}-p)^{2}%
-A(p-k_{2})A(p-k_{1})\right]  .
\end{equation}
\bigskip The Fourier transform $A(p)$ is given as usual by%

\begin{equation}
A(\,p\,)\;=\;\frac{4\,\pi}{|\,\vec{p}\,|^{2}}\: .
\end{equation}
\newline The terms in Eq.~(\ref{eq:Ec}) have the following meaning after we
expand the $\ln$-terms. The first term represents the second order exchange
term, Fig.~\ref{exch2}. The term ${\mathcal{H}}(p)$ is responsible for
generating the ring diagrams, Fig.~\ref{ring}. The term ${\mathcal{M}}(p)$ is
responsible for generating some of the ring diagrams with exchange. The cross
terms give the remaining ring diagrams with exchange,
 Fig.~\ref{firstlog}. The $\Pi(p)$ term generates ring 
diagrams with a self-energy insertion in each
ring, fig.~\ref{ringSE}. Hence the terms in the first $ln$ should contribute
the most. Some of the terms that appear upon expansion of the
second $\ln$ term are shown in Fig.~\ref{secondlog}. The 
term ${\mathcal{K}}(p)$ generates
the so called `anomalous' diagrams, Fig.~\ref{anomalous}.\cite{rebei} \ The term
$\Sigma_{1}(p)$ generates ladder terms like those in Figs.~\ref{sigma1} and
~\ref{ladder3}. $\Sigma_{2}(p)$ generates terms like those in
Fig.~\ref{sigma2}.
Cross terms of the last two terms are shown in
Fig.~\ref{crossTerm}.
Diagrams that appear in Fig.~\ref{coul} are important in a nonhomogeneous
medium and must be accounted for since they no longer vanish. They arise
whenever  $\rho(x,x)$ is not equal to the ion density  in the 
 energy expression, Eq.(\ref{e59}).

\begin{figure}[ptb]
\vspace*{13pt}
\centerline{\vbox{\hrule width 5cm height 0.001pt}}
\vspace*{1.4truein}
\centerline{\vbox{\hrule width 5cm height 0.001pt}}
\vspace*{13pt}
\caption{Second order exchange
diagram in Eq.~\ref{eq:Ec}. }
\label{exch2}
\end{figure}

\begin{figure}[ptb]
\vspace*{13pt}
\centerline{\vbox{\hrule width 5cm height 0.001pt}}
\vspace*{1.4truein}
\centerline{\vbox{\hrule width 5cm height 0.001pt}}
\vspace*{13pt}
\caption{Ring diagrams generated
by ${\mathcal{H}}(p)$ in Eq.~(\ref{eq:Ec}). }%
\label{ring}%
\end{figure}

\begin{figure}[ptb]
\vspace*{13pt}
\centerline{\vbox{\hrule width 5cm height 0.001pt}}
\vspace*{1.4truein}
\centerline{\vbox{\hrule width 5cm height 0.001pt}}
\vspace*{13pt}
\caption{Diagrams that appear in the first ln term
in Eq.(~\ref{eq:Ec}). }%
\label{firstlog}%
\end{figure}

\begin{figure}[ptb]
\vspace*{13pt}
\centerline{\vbox{\hrule width 5cm height 0.001pt}}
\vspace*{1.4truein}
\centerline{\vbox{\hrule width 5cm height 0.001pt}}
\vspace*{13pt}
\caption{Ring diagrams
generated by $\Pi(p)$ in Eq.~(\ref{eq:Ec}). }%
\label{ringSE}%
\end{figure}

\begin{figure}[ptb]
\vspace*{13pt}
\centerline{\vbox{\hrule width 5cm height 0.001pt}}
\vspace*{1.4truein}
\centerline{\vbox{\hrule width 5cm height 0.001pt}}
\vspace*{13pt}
\caption{Diagrams that appear in the second ln term
in Eq.(~\ref{eq:Ec}).}%
\label{secondlog}%
\end{figure}

\begin{figure}[ptb]
\vspace*{13pt}
\centerline{\vbox{\hrule width 5cm height 0.001pt}}
\vspace*{1.4truein}
\centerline{\vbox{\hrule width 5cm height 0.001pt}}
\vspace*{13pt}
\caption{A new infinite set of
diagrams at nonzero temperature.}
\label{anomalous}
\end{figure}

\begin{figure}[ptb]
\vspace*{13pt}
\centerline{\vbox{\hrule width 5cm height 0.001pt}}
\vspace*{1.4truein}
\centerline{\vbox{\hrule width 5cm height 0.001pt}}
\vspace*{13pt}
\caption{Several terms that
appear due to $\Sigma_{2}$.}%
\label{sigma2}%
\end{figure}

\begin{figure}[ptb]
\vspace*{13pt}
\centerline{\vbox{\hrule width 5cm height 0.001pt}}
\vspace*{1.4truein}
\centerline{\vbox{\hrule width 5cm height 0.001pt}}
\vspace*{13pt}
\caption{A term that is
both due to $\Sigma_{1}$ and $\Sigma_{2}$. Here three particles are 
interacting pairwise with two of the three particles exchanging states.}%
\label{crossTerm}%
\end{figure}

\begin{figure}[ptb]
\vspace*{13pt}
\centerline{\vbox{\hrule width 5cm height 0.001pt}}
\vspace*{1.4truein}
\centerline{\vbox{\hrule width 5cm height 0.001pt}}
\vspace*{13pt}
\caption{Coulomb interactions
that must be taken into account in inhomogeneous media. These diagrams 
vanish in the homogeneous case.}%
\label{coul}%
\end{figure}
Normalizing the momentum with respect to $k_{F}$, the
Fermi momentum, the explicit expression for the correlation energy per
particle in Rydbergs (Ryd.) has the following form,

\begin{eqnarray}
 \lefteqn{ \frac{E_{c}}{N}   =\frac{3}{16\pi^{5}}
\int d^{3}q\,d^{3}p_{1}\,d^{3}
p_{2}\frac{\theta(1-p_{2})\theta(1-p_{1})
\theta(E_{q_{1}}-1)\theta(E_{q_{2}}-1) }{q^{2} (q+p_{_{1}}-p_{_{2}} )^{2}
( q^{2}+q\cdot p_{1}-q\cdot p_{2} ) } } \nonumber\\
& & + \frac{3}{4\pi\alpha^{2}r_{s}^{2}} \int_{-\infty}^{\infty}
dx\int_{0}^{\infty} dy\: y^{2} \left \{ \ln \left[ 
1+{\mathcal{H}} (x,y)+{\mathcal{M}}(x,y)
 - 2\Pi(x,y) \right ] \right. \nonumber\\
& &  \left. -{\mathcal{H}}(x,y)-
{\mathcal{M}}(x,y)+2\Pi(x,y) \right \} \nonumber\\
& & -\frac{3}{2\pi\alpha^{2}r_{s}^{2}} \!\int_{-\infty}^{\infty}\!dx\int
_{0}^{\infty}\!dy\:y^{2} \left \{ \ln \left [ 1+{\mathcal{K}}
(x,y)+\Sigma(x,y)\right ] -{\mathcal{K}}(x,y)-\Sigma(x,y) \right \},
\label{eq:Ecp}
\end{eqnarray}
where we have set $E_{q_{1}}=|\vec{q}+\vec{p}_{1}|$ and $E_{q_{2}}=|\vec
{q}-\vec{p}_{_{2}}|$.\newline For convenience, we set
\begin{equation}
\frac{E_{c}}{N}=I_{1}\,+\,I_{2}\,+\,I_{3}%
\end{equation}
\newline where $I_{1}$, $I_{2}$, and $I_{3}$ are the first, the second and the
third integrals, respectively. \ The functions ${\mathcal{K}}(x,y)$, 
${\mathcal{H}}(x,y)$ and
$\Sigma_{1}(x,y)$ are given by,
\begin{equation}
{\mathcal{K}}(x,y)\,=\,\frac{4\alpha r_{s}g(y)}{\pi(-ix+y^{2}-1)} ,\label{e89}
\end{equation}

\begin{equation}
{\mathcal{H}}(x,y)\,=\,-e^{2}\,A(x,y)D(x,y)
\end{equation}
 or more explicitly, we have
\begin{eqnarray}
{\mathcal{H}}(x,y)  & =& \frac{2\alpha r_{s}}{\pi y^{2}}\left[  1+\frac{1}{8y^{3}%
}\left(  x^{2}+4y^{2}(1-\frac{y}{2})(1+\frac{y}{2})\right)  \ln\left(
\frac{(\frac{x}{2})^{2}+y^{2}(1+\frac{y}{2})^{2}}{(\frac{x}{2})^{2}%
+y^{2}(1-\frac{y}{2})^{2}}\right)  \right. \nonumber\\
& & \left.  -\frac{x}{2y}\left(  \arctan\left(  \frac{y(\frac{y}{2}+1)}{\frac
{x}{2}}\right)  -\arctan\left(  \frac{y(\frac{y}{2}-1)}{\frac{x}{2}}\right)
\right)  \right]
\end{eqnarray}
 and

\begin{equation}
\Sigma_{1}(x,y)=\frac{16\alpha^{2}r_{s}^{2}}{(2\pi)^{4}}\!\int\!d^{3}%
k_{1}d^{3}k_{2}g(x,y,k_{1},k_{2})
\end{equation}
where
\begin{equation}
g(x,y,k_{1},k_{2})=\frac{1}{|\vec{y}-\vec{k}_{1}|^{2}}\frac{1}{|\vec{y}%
-\vec{k}_{2}|^{2}}\times\frac{\left(  \theta(1-k_{1})-\theta(1-k_{12})\right)
\left(  \theta(k_{1}-E^{\prime})-\theta(1-k_{2})\right)  }{(-ix+y^{2}%
-1)(ix-k_{2}^{2}-k_{1}^{2}+E^{\prime}-1)}
\end{equation}
$\theta(x)$ is the step function. $E^{\prime}$, $g(y)$ and $\alpha$
are given by
\begin{equation}
E^{\prime}=|\vec{k}_{1}+\vec{k}_{2}-\vec{y}|^{2},
\end{equation}%
\begin{equation}
\alpha=\left(  \frac{4}{9\pi}\right)  ^{\frac{1}{3}},
\end{equation}
and%

\begin{equation}
g(y)=1+\frac{1-y^{2}}{2y}\ln\left|  \frac{1+y}{1-y}\right| .
\end{equation}
\newline The Gell-Mann-Brueckner term can be isolated from the full expression
for the correlation, Eq.(~\ref{eq:Ecp}), by rewriting the second integral,
$I_{2} $, in the following way:%

\begin{equation}
I_{2}=I_{2}^{ring}+I_{2}^{exch}+I_{2}^{seRing}%
\end{equation}
where
\begin{eqnarray}
I_{2}^{ring}  & = &\frac{3}{4\pi}\frac{1}{r_{s}^{2}\alpha^{2}}\int_{-\infty
}^{+\infty}dx\int_{0}^{\infty}dy\,\: y^{2} \left \{ \ln \left [
1+{\mathcal{H}}(x,y) \right ] -{\mathcal{H}}(x,y) \right \} \:
\label{eq:ring}\\
I_{2}^{exch}  & = &\frac{3}{4\pi}\frac{1}{r_{s}^{2}\alpha^{2}}\int_{-\infty
}^{\infty}dx\int_{0}^{\infty}dy\: y^{2}\left \{  \ln \left[  1+\frac
{{\mathcal{M}}(x,y)}{1+{\mathcal{H}}(x,y)} \right]  -{\mathcal{M}}(x,y) \right\}.
\label{eq:exch}%
\end{eqnarray}
\newline $I_{2}^{seRing}$ is what is left of $I_{2}-I_{2}^{ring}-I_{2}^{exch}%
$. The term $I_{2}^{ring}$ is indeed the full expression for the RPA term
$\frac{1}{N}E_{c}^{RPA}.$ By taking the limit $r_{s}\rightarrow0$ and $y<1$,
it reduces to the Gell-Mann-Brueckner result  \cite{gellmann}

\begin{equation}
\frac{E_{c}^{RPA}}{N}\approx0.0622\ln r_{s}-0.142+\ldots
\end{equation}
\medskip as we show next. Going back to the expression for ${\mathcal{H}}(p)$,
Eq.(~\ref{eq:AD}), we have

\begin{equation}
I_{2}^{ring}=\frac{1}{2}\int\frac{d^{3}k}{(2\pi)^{3}}\int\frac{d\omega}{2\pi
} \left [ \ln \left (  1+{\mathcal{H}}(k,\omega) \right )  -{\mathcal{H}}
(k,\omega) \right ]
\end{equation}
\newline with ${\mathcal{H}}(k,\omega)$ defined to be
\begin{equation}
{\mathcal{H}}(k,\omega)=2e^{2}\frac{4\pi}{k^{2}}\!\int\!\frac{d^{3}p}{(2\pi
)^{3}}\,\frac{\left(  \frac{1}{2}(\vec{p}+\vec{k})^{2}-\frac{1}{2}\vec{p}%
^{2}\right)  }{\omega^{2}+\left(  \frac{1}{2}(\vec{p}+\vec{k})^{2}-\frac{1}%
{2}\vec{p}^{2}\right)  }\left(  \theta(\mu-\frac{p^{2}}{2})-\theta(\mu
-\frac{1}{2}(\vec{p}+\vec{k})^{2})\right) .
\end{equation}
\medskip Now we set
\begin{equation}
{\mathcal{H}}(k,\omega)=\frac{2e^{2}}{(2\pi)^{3}}\,\frac{4\pi}{k^{2}}
 \left ( H_{1}(k,\omega)+H_{2}(k,\omega) \right )
\end{equation}
\newline Terms other than $I_{2}^{ring}$ in Eq.(~\ref{eq:Ecp}) provide 
corrections to the RPA-term. The first term in Eq.(~\ref{eq:Ecp} ), as we
mentioned above, is the second order exchange term,\cite{onsager}

\begin{eqnarray}
E_{c}^{2exch}  & = & \frac{1}{3}\ln2-\frac{3}{2\pi^{2}}\zeta(3)\nonumber \\
& & \simeq0.04836\it{Ryd}.
\end{eqnarray}
\medskip To compare our expansion to others we calculate the term that is
equivalent to the RPA calculation. We start by evaluating the integral $H_{1}
$. After integrating the angular variables, we have%

\begin{equation}
H_{1}(\omega,k)=\pi\int_{0}^{k_{F}}\frac{p\,dp}{k}\,\ln\left[  \;\frac
{\omega^{2}+(pk+\frac{1}{2}k^{2})^{2}}{\omega^{2}+(-pk+\frac{1}{2}k^{2})^{2}%
}\;\right] .
\end{equation}
\newline This integral can be easily performed over p. We get after that%

\begin{eqnarray}
H_{1}(\omega,k)  & = &\frac{\pi}{k^{3}} \left \{  \: \frac{1}{2}
\ln \left [ \;\omega^{2}+k^{2}(k_{F}+\frac{k}{2})^{2} \right ]
 \left [ \omega^{2}+k^{2}(k_{F}-\frac{k}{2})(k_{F}+\frac{k}{2}) \right ]
 \right. \nonumber\\
& & - \frac{1}{2} \ln \left [ \omega^{2}+k^{2}(-k_{F}+ \frac
{ k}{2} )^{2} \right ] \, \left [ \omega^{2}-k^{2}(k_{F}
+\frac{k}{2})(-k_{F}+\frac{k}{2}) \right ] \nonumber\\
& & \left.  -k^{2} \omega \left \lbrack \arctan\left(  \frac{k(k_{F}
+\frac{ k}{2})}{\omega} \right)  -\arctan \left(  \frac{k(-k_{F}
+\frac{ k}{2})}{\omega}\right)  \right ] +k_{F}k^{3} \right\}  .
\end{eqnarray}
\newline The second integral $H_{2}$ in ${\mathcal{H}}(\omega,k)$ is seen to be
simply obtained from $H_{1}(\omega,k)$ by replacing $\vec{k}$ by $-\vec{k}$.
Therefore we have%

\begin{equation}
{\mathcal{H}}(\omega,k)\, = \, 2 H_{1}( \omega, k )
\end{equation}
\newline If we set ${y=\frac{k}{k_{F}}}$ and ${x=\frac{\omega}{\mu}}$, we get%

\begin{eqnarray}
{\mathcal{H}}(x,y)  & = & \frac{2\alpha r_{s}}{\pi y^{2}}\left\{  1+\frac
{x^{2}+4y^{2}(1-\frac{y^{2}}{4})}{(2y)^{3}}\ln\left[  \frac{x^{2}
+4y^{2}(1+\frac{ y}{2})^{2}}{x^{2}+4y^{2}(1-\frac{ y}{2})^{2}
}\right]  \right. \nonumber\\
& & \mbox{}\left.  -\frac{x}{2y}\left[  \arctan\left(  \frac{2y(1+\frac
{ y}{2})}{x}\right)  +\arctan\left(  \frac{2y(1-\frac{ y}{2}%
)}{x}\right)  \right]  \right\}
\end{eqnarray}
\newline where $r_{s}=\frac{e^{2}}{\alpha\,k_{F}}$. Hence the ring diagrams'
contribution to the energy is given by%

\begin{equation}
E_{c}^{RPA}=\frac{3N}{4\pi\alpha^{2}r_{s}^{2}}\int_{0}^{\infty}y^{2}%
\,dy\int_{0}^{\infty}dx\left[  \ln[\,1+{\mathcal{H}}(x,y)]-{\mathcal{H}}(x,y)\right]  ,
\end{equation}
\newline where $N$ is the total number of electrons. Now, we notice that if we
make the following substitutions:%

\begin{eqnarray}
y\,  & = & \,y^{\prime}\nonumber\\
x\,  & = & \,2yx^{\prime}%
\end{eqnarray}
\medskip we recover exactly the same expression for the ring diagrams as
obtained by Bishop and Luhrmann (BL) using a totally different
expansion. \cite{bishop} \ This helps to validate our original expansion in $\hbar$ and later
in iterating the equations of motion in terms of $e^{2}$. We stress that this
expansion is valid for both small and large $r_{s}$.

From the above analysis, we see explicitly that the method of Effective Action
amounts to including another infinite set of diagrams besides the usual ring
diagrams. One subset of the diagrams added is the ring diagrams that allow
exchange in them, 
Fig.~\ref{firstlog}. On physical grounds these are
expected to be the next important ones that must be summed up. It is also
easily seen from the above that all second order diagrams are included in the
expansion with the right symmetry factors. Once more, this shows 
that our original
expansion in $\hbar$ is indeed meaningful. One last thing to note about the
diagrams in Fig.~\ref{secondlog} is that they include the ``anomalous''
diagrams which appear due to the finite temperature 
method. \cite{luttinger}\  From a nonzero temperature 
calculation, we were able to show
that these diagrams give a zero contribution at zero temperature simply
because of the  Fermi statistics. \cite{rebei} From 
Eq.(\ref{eq:Ecp}) the contribution of 
these diagrams involves integrating $\ln \left [
1 + {\mathcal{K}}(x,y) \right ] - {\mathcal{K}}(x,y)$ over all $x$ in the 
complex plane. Since ${\mathcal{K}}(x,y)$ has a simple pole, Eq.(\ref{e89}), the above 
integrand ends up having poles of order two and higher, resulting
in  their zero
contribution at zero temperature. However these diagrams become essential at
non-zero temperature where the above constraint of Fermi statistics is 
no longer an issue. In the following section  we calculate 
the contribution of the function
${\mathcal{M}}(x,y)$ to the correlation energy.

\begin{figure}[ptb]
\vspace*{13pt}
\centerline{\vbox{\hrule width 5cm height 0.001pt}}
\vspace*{1.4truein}
\centerline{\vbox{\hrule width 5cm height 0.001pt}}
\vspace*{13pt}
\caption{Ring diagrams
with exchange.}%
\label{RingExchange}%
\end{figure}

\begin{figure}
\vspace*{13pt}
\centerline{\vbox{\hrule width 5cm height 0.001pt}}
\vspace*{1.4truein}
\centerline{\vbox{\hrule width 5cm height 0.001pt}}
\vspace*{13pt}
\caption{Comparison of the Hubbard approximation and the BL approximation to 
the Coulomb exchange interaction : 
$ \langle \frac{1}{|\vec{k}_1 -\vec{k}_2 + \vec{y}|^2} \rangle$ vs. 
$y$ .}
\label{HBL_graph}
\end{figure}

\begin{figure}[ptb]
\vspace*{13pt}
\centerline{\vbox{\hrule width 5cm height 0.001pt}}
\vspace*{1.4truein}
\centerline{\vbox{\hrule width 5cm height 0.001pt}}
\vspace*{13pt}
\caption{ Comparison of final results for the correlation energy (in Ryd.) of 
a homogeneous Fermi gas:  1. Ref.\cite{ceperly}, 
2. Ref. \cite{bishop},  3. This paper }
\label{table}
\end{figure}

\begin{figure}[ptb]
\vspace*{13pt}
\centerline{\vbox{\hrule width 5cm height 0.001pt}}
\vspace*{1.4truein}
\centerline{\vbox{\hrule width 5cm height 0.001pt}}
\vspace*{13pt}
\caption{ Diagrams that appear when we include terms of order $\hbar^{6}$ in
$\Gamma$. }
\label{nextTerm}
\end{figure}

\section{The Inclusion of Second Order Exchange Effects \\ in Ring Diagrams}
\noindent
In this section, we give the contribution of diagrams like those shown in
Fig.~\ref{RingExchange} to the correlation energy at zero temperature. 
Hence
we go beyond RPA in this case. We will show that our method provides excellent
agreement with fully numerical calculations. We also compare our results to
those found by BL.\cite{bishop} where a coupled cluster formalism has 
been used to get the correlation energy. To get to these final results we had to make approximations
along the way. We use two different approximations: the Hubbard approximation
\cite{hubbard} and the Bishop-Luhrmann 
approximation.\cite{bishop} We have
found that the former applies well to high values of $r_{s}$ while the latter
applies well to low values of $r_{s}$. From Eq.(\ref{eq:exch}), this amounts to finding
$I_{2}^{exch}$ and the second order exchange term that has already been
calculated by Onsager et al.\cite{onsager} \ Now we 
show how to calculate
$I_{2}^{exch}$. The calculation is straightforward but special care must be
exercised when it comes to numerically evaluating the final result. First we 
let

\begin{equation}
I=\int_{-\infty}^{\infty}dx\int_{0}^{\infty}y^{2}\;dy\ln\left[  1+\frac
{{\mathcal{M}}(x,y)}{1+{\mathcal{H}}(x,y)}\right]
\end{equation}
where

\begin{equation}
{\mathcal{M}}(x,y)= \frac{16\alpha^{2}r_{s}^{2}}{(2\pi)^{4}}\frac{1}{y^{2}}
\int_{\Gamma}\frac{d^{3}k_{1}d^{3}k_{2}}{|\vec{k_{1}}-\vec{k_{2}}+\vec{y}
|^{2}} \frac{1}{ \left ( -ix+ \frac{\left ( (\vec{k_{1}}+\vec{y})^{2}
-k_{1}^{2} \right )}{2} \right ) \left ( ix+\frac{\left ( (\vec{k_{2}} -
\vec{y} )^{2} - k_{2}^{2} \right ) }{2} \right )} 
\end{equation}
and where the region of integration $\Gamma$ is given by:

\begin{equation}
\Gamma= \left \{ (k_{1},k_{2})k_{1}<1,\: k_{2}<1,|\vec{y}+\vec{k_{1}
}|>1,\: |\vec{y}-\vec{k_{2}}|>1 \right \} \: .
\end{equation}

It is obvious from the above that we are faced with a daunting task of having
to deal with a 9-D integral inside a logarithmic function which itself 
has to be
integrated over normalized energy $x$ and normalized momentum $y$. Most of the
complications are related to the angle integrals which can not be separated.
To be able to make some progress we have to make an approximation, i.e., we
either assume that $|\vec{k}_{i}+\vec{y}|\approx1$ and average over the angle
between them, i.e., the Hubbard approximation or we use a more sophisticated
approximation like the one proposed by Bishop and Luhrmann (BL). Hubbard's
approximation amounts to the following:
\begin{equation}
\frac{1}{|\vec{k_{1}}-\vec{k_{2}}+\vec{y}|^{2}}\approx\frac{1}{y^{2}+1}.
\end{equation}
On the other hand, the BL-approximation is more complicated and the reader is
referred to their appendix \cite{bishop}  for 
a discussion of their approximation. \ Both 
approximations are almost equivalent for large values
of momentum $y$, i.e. $y>2$ in units of Fermi momentum. For values of $y$ less
than 2, both approximations
are quite different ( see Fig.~\ref{HBL_graph} ). The Hubbard 
approximation seems to apply well for moderately large values of $r_{s}$ while
the BL-approximation applies well for low values of $r_{s}$. In fact the 
BL-approximation incorporates the Fermi statistics of the interacting
particles where it matters most, i.e. in high density situations,
and this is one 
reason why we found it to be a good approximation to the exchange Coulomb
interaction for low $r_{s}$. The Hubbard approximation should be expected to
be a good approximation for particles near the Fermi surface and where 
statistics
are not an issue. This case  should apply well to
 low density.
Given either approximation we end up with an
expression for ${\mathcal{M}}(x,p)$ of the form:
\begin{equation}
{\mathcal{M}}(x,y)=\frac{16\alpha^{2}r_{s}^{2}}{(2\pi)^{4}}\frac{1}{y^{2}
}f(y)\int_{\Gamma}d^{3}k_{1}d^{3}k_{2}\frac{x^{2}+\frac{1}{4}(y^{2}
+2\vec{k_{1}}\cdot\vec{y})(y^{2}-2\vec{k_{2}}\cdot\vec{y})}{\left (
x^{2}+\frac{1}{4}(y^{2}+2\vec{k_{1}}\cdot\vec{y})^{2} \right ) \left (x^{2}+
\frac{1}{4}(y^{2}-2\vec{k_{2}}\cdot\vec{y})^{2} \right )}
\end{equation}
where the function $f(y)$ is our approximation to the exchange term in
${\mathcal{M}}(x,y)$ and is assumed known. \ 
The integrations over $\vec{k_{1}}$ and $\vec{k_{2}}$ are easily
performed  if we
choose to write both vectors in cylindrical coordinates with $\vec{y}$ along
the z-axis, i.e.,
\begin{equation}
\vec{k_{i}}=(\rho_{i},z_{i},\theta_{i}),\: \: \: \: \:  i=1,2 \:.
\end{equation}
Hence the integration over $\vec{k_{1}}$ becomes:
\begin{equation}
\int d^{3}k_{1}\;=\;\int_{0}^{2\pi}\;d\theta_{1}\;\int_{\Gamma^{\prime}}%
\rho_{1}\;d\rho_{1}dz_{1},
\end{equation}
where
\begin{eqnarray}
\Gamma^{\prime}&=&\left \{ y < 2,\: \: \: -\frac{y}{2} < z_{1} < 1-y,\: \: \:
\left ( 1-(z_{1}+y)^{2} \right ) ^\frac{1}{2} < \rho_{1} < (1-{z_{1}}
^{2})^{^\frac{1}{2}} \right \}\nonumber \\
&& \bigcup \left \{ y<2,\: \: \: 1-y < z_{1} < 1,\: \: \: 0 < \rho_{1} < 
(1-{z_{1}}^{2})^{^\frac{1}{2}} \right \}\nonumber \\
&& \bigcup \left \{ y > 2,\: \: \: -1 < z_{1} < 1,\: \: 0 < \rho_{1} <
(1-{z_{1}}^{2})^{\frac{1}{2}} \right \},
\end{eqnarray}
with an equivalent expression for the region of integration over $k_{2}$. The
integrations are now easily carried out. The expression that we get for
${\mathcal{M}}(x,y)$ is naturally expressed in terms of $\frac{2x}{y}$ instead
of $x$, so in the following $x$ refers to $\frac{2x}{y}$. We only quote the
final expression here:%

\begin{equation}
{\mathcal{M}}(x,y)=-16\frac{\alpha^{2}{r_{s}}^{2}}{\pi^{2}}\frac{1}{y^{4}%
}f(y){\mathcal{L}}(x,y),
\end{equation}
with
\begin{eqnarray}
{\mathcal{L}}(x,y)  & = &\left(  -\frac{y^{2}}{16} \left ( \frac{\pi y^{2}
}{2\alpha r_{s}} \right )^{2}{\mathcal{H}}(x,y) \left ({\mathcal{H}}
(x,y)-2 \right ) \right.  \nonumber\\
& & \mbox{}-\frac{(yx)^{2}}{64}\left\{  y+ \left ( \frac{x}{2}+\frac
{2(1-\frac{y^{2}}{4})}{xy} \right )\left ( \arctan \left ( \frac{2-y}{x}
\right ) - \arctan \left ( \frac{2+y}{x} \right ) \right ) \right.
\nonumber\\
& & \mbox{}\left.  +\frac{1}{2}\ln \left[  \frac{(x^{2}+(2-y)^{2})}
{(x^{2}+(2+p)^{2})}\right]  \right \}  \nonumber\\
& & \times\left\{  -1+\frac{x}{2} \left ( 1-\frac{y^{2}}{4}+\frac{x^{2}}
{4} \right )\left( \arctan \left ( \frac{2-y}{x} \right ) - \arctan \left (
\frac{2+y}{x} \right ) \right ) + \right.  \nonumber\\
& & \left.  \left.  \ln \left[  \frac{ \left ( x^{2}+(2-y)^{2} \right )
^{\frac{1}{2}} \left ( x^{2}+(2+y)^{2} \right )^{\frac{1}{2}}}{x^{2}
}\right]  \right\}  \right)  \Theta(2-y)\nonumber\\
& & \mbox{}+\left(  -\frac{y^{2}}{64}(\frac{\pi y^{2}}{2\alpha r_{s}}
)^{2}{\mathcal{H}}(x,y)({\mathcal{H}}(x,y)-1)\right.  \nonumber\\
& & \mbox{}-\frac{yx^{2}}{128}\left\{  -4-y\ln\left[  \frac{\left (
x^{2}+(2-y)^{2} \right )^{\frac{1}{2}} \left ( x^{2}+(2+y) \right )
^{\frac{1}{2}}}{x^{2}}\right]  \right.  \nonumber\\
& & \mbox{}+\left.  \frac{4}{x} \left ( 1-\frac{y^{2}}{4}+\frac{x^{2}}
{4})(\arctan(\frac{2+y}{x})-\arctan(\frac{-2+y}{x}) \right ) \right\}
\nonumber\\
& & \times\left\{  \frac{1}{2}\ln\left[  \frac{ \left ( x^{2}+(2-y)^{2}
\right )^{\frac{1}{2}}\left (x^{2}+(2+y)\right )^{\frac{1}{2}}}{x^{2}
}\right]  \right.  \nonumber\\
& & \mbox{}+\left.  \left.  \left.  \frac{2}{x}\frac{1-\frac{y^{2}}{4}}
{y}\left (\arctan(\frac{-2+y}{x})-\arctan(\frac{2+y}{x})\right )\right\}
\right\}  \right)  \Theta(y-2) \: .
\end{eqnarray}
Given this expression for ${\mathcal{M}}(x,y)$, we can easily write the full
expression for the correlation due to exchange effects in ring diagrams. It is
important to note that we allow for more than one exchange to take place in
each ring diagram. Hence our original integral $I$, can be now evaluated
numerically. So within the above approximation, we were able to reduce our
calculations to a two-dimensional integral:
\begin{equation}
I_{2}^{exch}=\frac{3}{4\pi}\frac{1}{(\alpha r_{s})^{2}}\int_{0}^{\infty}%
dx\int_{0}^{\infty}y^{3}\left(  \ln\left[  1+\frac{{\mathcal{M}}(x,y)}%
{1+{\mathcal{H}}(x,y)}\right]  -{\mathcal{M}}(x,y)\right) .
\end{equation}

The expression for the function $f(p)$ depends on which approximation we
choose to use. In this formalism it is hard to decide on which one since both
give very close answers to the second order exchange diagram. We have found
that the Hubbard approximation in this instance gives 0.041 $Ryd.$ while the BL
approximation gives 0.049 $Ryd$. Clearly the latter is closer to the true value
of 0.0484 $Ryd.$, but we believe that this is not enough to decide against using the
Hubbard approximation. In fact at low $r_{s}$ we expect that the high energy
part of the exchange energy between particles to dominate and this explains
why we found the BL-approximations to apply well in this range. At high
$r_{s}$ this is not true anymore and in this instance we expect the
BL-approximation to overestimate this exchange of energy between the
particles. This is in fact what we have found. This creates doubt that the
BL-approximation is always better than the Hubbard approximation. As we 
stated 
above, the second order exchange term is not a good test since it is density
independent. This explains why the Hubbard approximation gives better results
for higher values of $r_{s}$. Below we give some numerical answers to compare
with those obtained using quantum Monte Carlo methods (QMC). \cite{ceperly}
Our results are obtained using both approximations. For the second order
exchange diagram we have used the true value of 0.0484 Ryd.

 Clearly our results, Fig.\ref{table}, agree 
well with previous results. For large
values of $r_{s},$ i.e., $r_{s}$ $\geq 10,$ our results do not compare well
to the QMC results. However this is expected since we need to include
corrections due to the ladder diagrams which are believed to be important at
low densities. This concludes our calculations. \ We have shown the
effectiveness of an Effective Action approach to electronic systems. In the
next section we shall examine the viability of using this formalism in the
nonhomogeneous case and show how it is related to density functional theory.

\section{ The Nonhomogeneous Electron System}
\noindent
In this last section, we  go back to the inhomogeneous case and
show how our method gives some of the results known in DFT. \cite{kohn}
$,$\cite{sham}\ We start by briefly reviewing the basic ideas behind DFT. The
Hamiltonian H is written as the sum of three terms: T, V and U. T is the
kinetic energy term, V is the external potential term and U is the Coulomb
energy term, so we have
\begin{equation}
H=T+V+U  \: \:,
\end{equation}%
\begin{equation}
T=\frac{1}{2}\int\nabla\psi^{\dagger}(r)\nabla\psi(r)d^{3}r,
\end{equation}%
\begin{equation}
V=\int V(r)\psi^{\dagger}(r)\psi(r)d^{3}r,
\end{equation}%
\begin{equation}
U=\frac{1}{2}\int\frac{\psi^{\dagger}(r)\psi^{\dagger}(r^{\prime}%
)\psi(r^{\prime})\psi(r)}{|r-r^{\prime}|}dr\;dr^{\prime}.
\end{equation}
\newline If $\Psi$ is the ground state, then the density $n(r)$ is given by
\begin{equation}
n(r)=\langle\Psi|\psi^{\dagger}(r)\psi(r)|\Psi\rangle.
\end{equation}
The first important fact to note is that $V(r)$ is a unique functional of
$n(r)$, assuming that there is a unique ground state for the system. Now let
the functionals $F[n(r)]$ and $E_{V}[n(r)]$ be defined as follows:%

\begin{equation}
F[n(r)]=\langle\Psi|T+U|\Psi\rangle,
\end{equation}%
\begin{equation}
E_{V}[n(r)]=\int V(r)n(r)d^{3}r+\;F[n(r)].
\end{equation}
Then, if we assume that there is a one-to-one correspondence between $n(r)$
and $V(r)$, it can be shown that
\begin{equation}
\frac{\delta E_{V}[n]}{\delta n(r)}|_{g.s}=0,
\end{equation}
with
\[
\int n(r)\;d^{3}r=N\: \: .
\]
\newline This result is called the Hohenberg-Kohn theorem. Next, the Coulomb
energy is isolated from the functional $F[n]$ by introducing a new functional
$G\left[  n\right]  $:
\begin{equation}
F[n]=\frac{1}{2}\int \: d\vec{r}\: d{\vec{r}}^{\prime}
\frac{n(r)\;n(r^{\prime})}{|r-r^{\prime}|}\: + \: G[n]\: ,
\end{equation}%
\begin{equation}
G[n]=\frac{1}{2}\int \:d\vec{r}\: \: \nabla_{r}\nabla_{r^{\prime}}
\: n_{1}(r,r^{\prime
})|_{r=r^{\prime}}+\frac{1}{2}\int\frac{C_{2}(r,r^{\prime})}{|r-r^{\prime}%
|}dr\;dr^{\prime},
\end{equation}
and
\begin{equation}
C_{2}(r,r^{\prime})=n_{2}(r,r^{\prime};r,r^{\prime})-n(r)n(r^{\prime}).
\end{equation}
\newline Here $n_{1}(r,r^{\prime})$ is the one-particle density matrix and
$C_{2}(r,r^{\prime})$ is a correlation function defined in terms of the one-
and two-particle density matrices. DFT calculations are essentially centered
around finding good approximations to the functional $G[n]$. This is usually
done by postulating that there is a virtual system of free electrons in an 
external potential with exactly the same density as the interacting
system. The energy is found by finding the eigenfunctions that correspond
to this external potential self consistently. 

  The
method that was presented makes use of concepts which are 
 similar in many ways to the 
ideas expressed in DFT. However there are major differences. Our method 
clearly incorporates the
Hohenberg-Kohn theorem. From Eq.(\ref{e35}), we have the following
\begin{eqnarray}
\frac{\delta E_{gs} \left [ \rho \right ]}{\delta \rho (x,x)} = - Q(x,x) \: . 
\end{eqnarray}
From Eq.(\ref{e49}), the term
\begin{eqnarray}
\int d^{3}x \: V(x) \rho(x,x) \: ,
\end{eqnarray}
can be separated from the energy $E_{gs}$. Hence  if we set $Q(x,x) = 0$ , we
immediately 
get the Hohenberg-Kohn result. In fact we now have  an explicit
expression for $E_{V}[n]$ within perturbation theory. This result also 
shows the one-to-one correspondence between density and external potential
as long as there is a unique solution around $(J,B,Q)=(0,0,0)$ for the 
defining equations, Eqs.(\ref{e33}-\ref{e35}). 

We have 
solved these equations only
approximately. \ We have made two important approximations. The first 
corresponds to the number of diagrams included in $\Gamma$ from the outset.
However from the calculations of the homogeneous case, we see that it 
will probably be the case that including only two diagrams in the inhomogeneous
case will give good results. \ Including higher order corrections by taking
into account diagrams like the ones in Fig.~\ref{nextTerm} is nevertheless
straightforward if more accurate results are desired. The second approximation
we have made was in solving for $\rho(x,y)$. We had to linearize 
the equations of motions to be able to solve for $\rho(x,y)$ iteratively. Also
from Eq.(\ref{e50}), we get the
following equivalent result:
\begin{equation}
\nabla^{2}\varphi_{c}(r)=4\pi n(r).
\end{equation}

\bigskip Clearly the method presented here 
is a generalization of the Thomas-Fermi 
method  where
$n(r)$ in the last equation is the \emph{true} density of the system. A very
important difference from DFT is that
within this  method we have a systematic scheme for calculating the functional
$G[n]$, Eq.(\ref{e49}). From the above analysis in the homogeneous case, it is 
obvious that the most important contribution to correlation at zero 
temperature comes from calculating the following term:
\begin{eqnarray}
Tr \ln {\mathcal{X}} = \ln \: det \frac{A^{-1} + e^{2} \rho \rho {\mathcal{X}} 
+ e^{4} {\mathcal{X}} {\mathcal{X}} \rho \rho A \rho \rho}{A^{-1}} \: 
\end{eqnarray}
where
\begin{eqnarray}
{\mathcal{X} } \: \equiv \: A^{-1} \: C \: .
\end{eqnarray}
To calculate the determinant of the above operators, we need a basis of
wavefunctions. In practical computations, it is more advantageous to use 
a finite basis set. If we choose as our basis set the wavefunctions
$\left \{ \phi_{i} \right \}_{i=1}^{N}$ with eigenvalues $\epsilon_{i}$
such that
\begin{eqnarray}
\left (-\frac{1}{2} \nabla^{2} + V(x) \right ) \phi_{i}(x) = \epsilon_{i} \phi_{i}(x),
\end{eqnarray}
then the wavefunctions, $\psi_{j}$, of the full interacting system 
of $N$ electrons can 
be written as a linear combination of the above wavefunctions:
\begin{eqnarray}
\psi_{j}(x,t) \: = \: \sum_{i=1}^{N} a_{ji} \phi_{i}(x) e^{-\epsilon_{i} t} \: .
\end{eqnarray}
The coefficients $a_{ji}$ are found self consistently 
by solving Eq.(\ref{e56}). To 
solve for ${\mathcal{X}}(x,y)$, we solve Eq.(\ref{e53}) 
self-consistently. Clearly 
this method is computationally more intensive than DFT. \ However 
approximations like the static approximation used by Kotani \cite{kotani} 
can simplify the calculations a lot. 

 Finally, we 
would like to point out that this method applies equally well to excited
states other than the ground state. The Hartree field of the excited state
is found again by using Eq.(\ref{e33}). If $\varphi_{0}(x)$ is the ground state
solution then $\varphi_{1}(x) = \varphi_{c}(x) + \Delta(x)$ is the 
excited state solution iff
\begin{eqnarray}
\frac{\delta \Gamma}{ \delta \varphi(x)} {\left | \right .}_{\varphi 
= \varphi_{1}} = 0.
\end{eqnarray}
This implies that
\begin{eqnarray}
\int d^{4}y \: \frac{\delta^{2} \Gamma}
{\delta \varphi(x) \delta \varphi(y)} {\left | \right .}_{
\varphi_{0}} \Delta(y) = 0.
\end{eqnarray}
In the above, $\Delta(x)$ is assumed small compared to $\varphi_{0}(x)$. This 
latter equation enables us to solve for $\Delta(x)$ and hence the corresponding
$\rho(x,y)$. \  We expect this 
method to apply equally well to atomic and molecular systems as we have shown
to be the case for homogeneous systems.

\bigskip
\bigskip

\section{Conclusion}
\noindent
We have used a functional method, the Effective Action method, to calculate 
correlation effects in electronic systems. \ A general 
expression for the energy 
was derived which includes corrections beyond the RPA approximation. \ We have 
also shown how to apply this result to a homogeneous electron gas. \ Our 
results agree well with quantum Monte Carlo calculations. \ The inclusion
of higher order corrections showed that the energy expression derived is 
not necessarily applicable only to high density cases. \ This method has 
many similarities with the principles on which Density Functional Theory
is founded. \ The main result is that a systematic expression for the 
correlation energy can be written down. \ The use of such expressions
in realistic calculations,  with 
approximations like those used by Kotani \cite{kotani}, should result 
in our being able to simplify the calculations, not only in the homogeneous 
case but also for the nonhomogeneous case. \

\section*{Acknowledgements}
\noindent
We thank C.J. Goebel for very useful discussions and L. Benkhemis for comments.

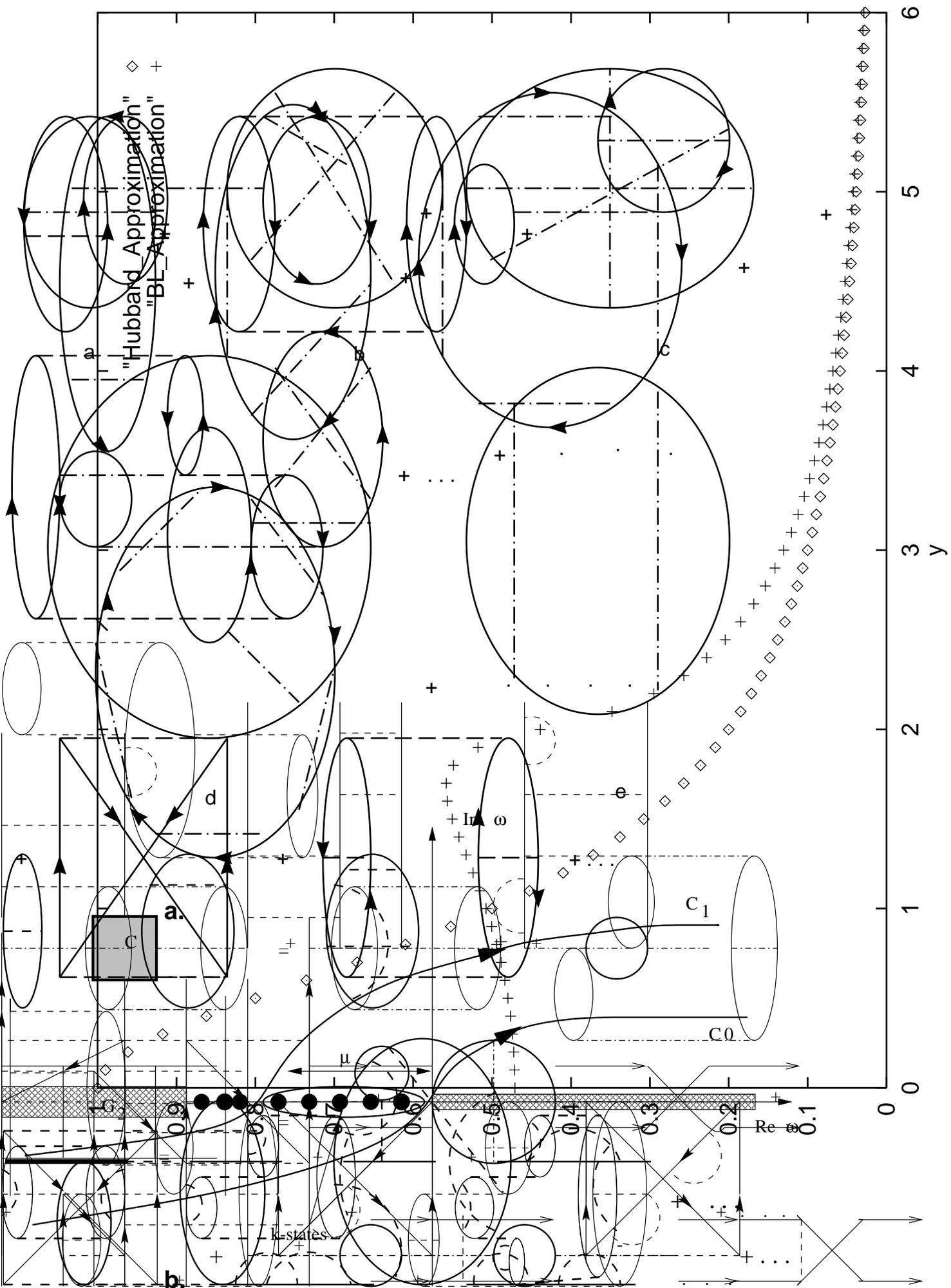
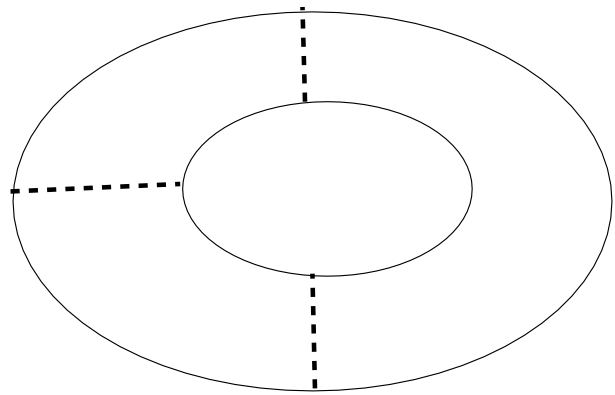
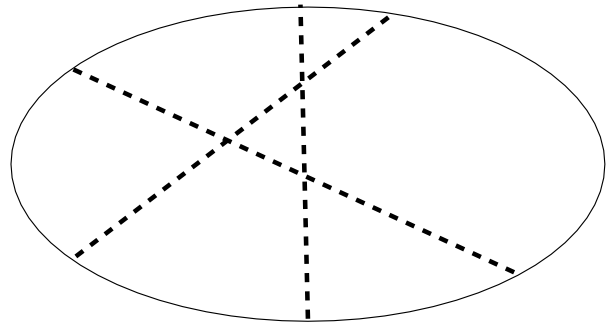
\end{document}